\documentclass[aps,pra,twocolumn,amsmath,amssymb,superscriptaddress,floatfix]{revtex4-1}

\usepackage{amsmath}
\usepackage{graphicx}
\usepackage{multirow}
\usepackage{color}
\usepackage{bm}

\usepackage[whole]{bxcjkjatype}

\def\t#1{\textrm{#1}}
\def\ket#1{|#1\rangle }

\def\braket#1{\langle #1 \rangle}
\def\n{\nonumber \\ }

\begin{document}

\title{Electric polarization and nonlinear optical effects in noncentrosymmetric magnets}

\author{Takahiro~Morimoto}
\thanks{These authors contributed equally.}
\affiliation{Department of Applied Physics, The University of Tokyo, Hongo, Tokyo, 113-8656, Japan}
\affiliation{JST, PRESTO, Kawaguchi, Saitama, 332-0012, Japan}

\author{Sota~Kitamura}
\thanks{These authors contributed equally.}
\affiliation{Department of Applied Physics, The University of Tokyo, Hongo, Tokyo, 113-8656, Japan}

\author{Shun~Okumura}
\affiliation{Institute for Solid State Physics, The University of Tokyo, Kashiwa 277-8581, Japan}

\begin{abstract}
We study electric polarization and nonlinear optical effects in spin systems with broken inversion symmetry.
We apply strong coupling expansion to the underlying electronic Hamiltonians, and systematically derive expressions for electric polarization in spin systems that are represented in terms of spin operators. The magnon representation of the obtained electric polarization operator allows us to compute linear and nonlinear optical responses by the standard diagrammatic method. We apply our formalism to Heisenberg model with alternating coupling constants and $J_1$-$J_2$ model with inversion symmetry breaking. We demonstrate that these inversion broken spin systems support dc current flow upon magnon excitations which arises from the shift current mechanism.
\end{abstract}

\date{\today}

\maketitle

\section{Introduction}

Nonlinear responses of quantum materials are actively studied due to both fundamental and technological importance \cite{Boyd,Bloembergen,Sturman}. For example, quantum materials with broken inversion symmetry exhibit photovoltaic effects for various intrinsic mechanisms including shift current \cite{Baltz-Kraut,Belinicher82,Sipe,Young-Rappe,Young-Zheng-Rappe,Cook17,Morimoto-Nagaosa16,Tan16,Sotome19,Burger19,Hatada20}, injection current \cite{Sipe,deJuan17,Orenstein21} and ballistic current \cite{Belinicher-Sturman80,Belinicher-Sturman88}, which suggests their potential application to solar cells and photodetectors. 
In particular, shift current is a photovoltaic effect in noncentrosymmetric crystals and arises from a geometrical origin. Specifically, the center of electron wave packet is shifted upon optical excitation and this motion of electrons leads to dc current response. 
The shift of the wavepacket is quantified by the so-called shift vector that is formulated with Berry connection. The shift current is closely related to the modern theory of polarization since the electric polarization is given by the Berry phase that is the integral of Berry connection over the Brillouin zone \cite{Resta,Kingsmith93,Vanderbilt93}.

Strongly correlated systems are known to support interesting optical properties \cite{Imada-RMP}. In particular, collective modes appear in low energy region, and they can be frequently accessible with optical excitation \cite{Tokura14,Shimano20}. However, previous studies on shift current responses have been mostly focused on systems of noninteracting electrons, and shift current from strongly correlated systems has been not fully explored so far. Since the low energy excitations can lead to larger coupling to electromagnetic fields via larger vector potential, the collective modes are expected to show an enhancement in nonlinear responses. Also, understanding of their nonlinear responses can reveal novel nonlinear functionalities of quantum materials with strong electron correlation. 

\begin{figure}
    \centering
    \includegraphics[width=0.9\linewidth]{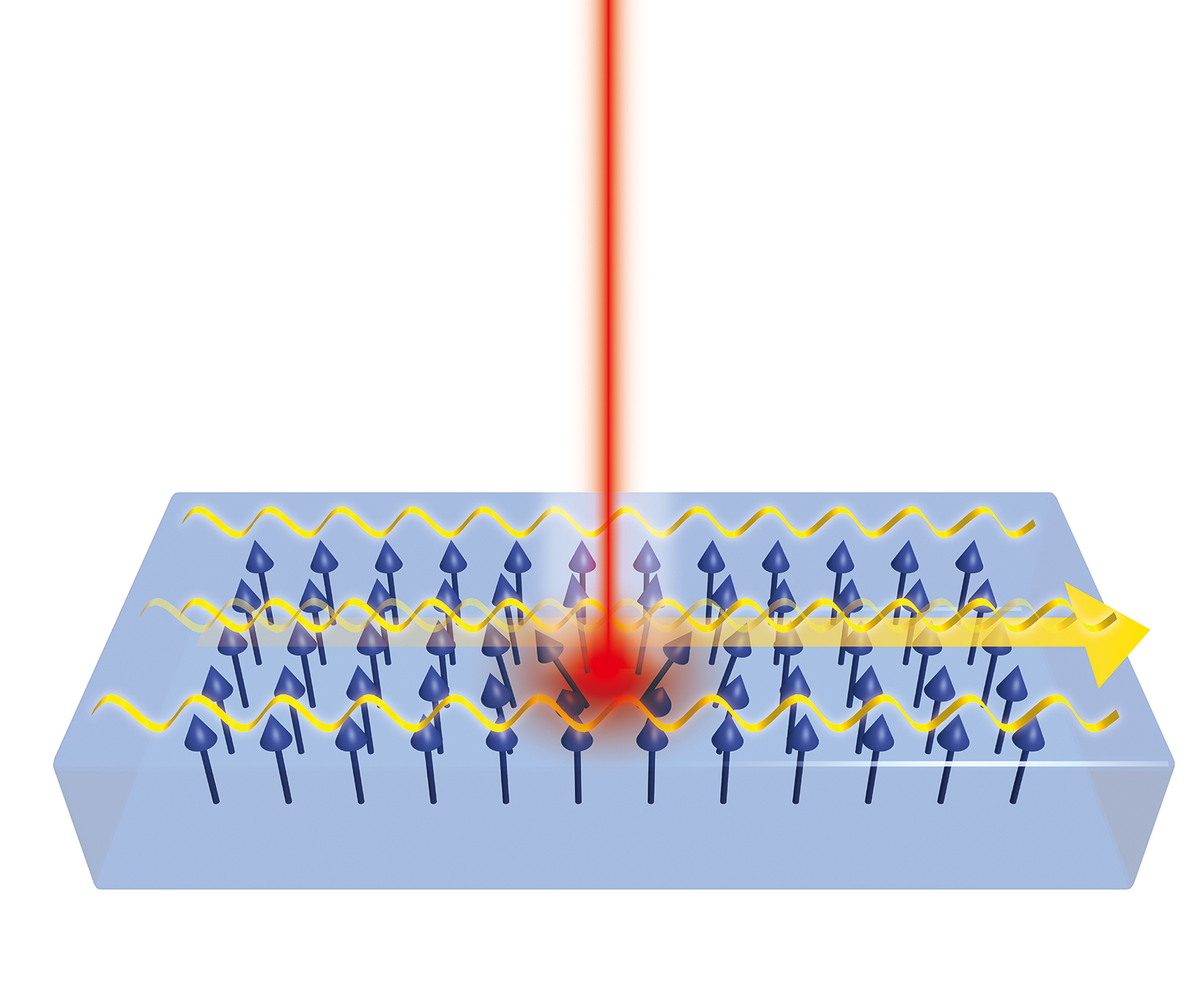}
    \caption{Schematic picture of shift current induced by magnon excitations under the light irradiation.}
    \label{fig:schematics}
\end{figure}

Among a variety of strongly correlated materials, here we focus on noncentrosymmetric magnets that exhibit multiferroic responses \cite{Kimura03,Katsura05,Mostovoy06,Tokura14}. The collective mode in noncentrosymmetric magnets is an electromagnon, which is a magnetic excitation that accompanies electric polarization. The nonzero polarization of magnons is closely related to the appearance of the electric polarization in the ground state of the noncentrosymmetric spin systems and is a consequence of their multiferroic nature \cite{Katsura05,Katsura07,Jia06,Jia07,Katsura09}. Electromagnons can be excited with external light field \cite{Pimenov06,Aguilar09,Takahashi12} and are shown to induce dichroism \cite{Kibayashi14,Kida}. 
Shift current response in noncentrosymmetric spin systems has been previously studied based on electronic models in the presence of spin-orbit coupling, by one of the authors  \cite{Morimoto-magnon19}. Specifically, as the electromagnons are excited by light irradiation, the polarization $P$ increases in time, which induces flow of dc current $J$ due to the relationship $J=dP/dt$ as schematically illustrated in Fig.~\ref{fig:schematics}.  In Ref.~\cite{Morimoto-magnon19}, shift current of magnons were demonstrated based on a 1D toy model, while its derivation strongly relies on the underlying electronic Hamiltonian and was difficult to apply to general spin systems described with spin Hamiltonians. Also, the magnitude of the obtained shift current was limited since it is proportional to the magnitude of spin-orbit coupling which is generally small. To fully explore shift current responses in general spin systems to seek large nonlinear functionality, it is highly desired to establish a more general framework that enables to study optical responses in terms of spin operators (without relying on the underlying electronic operators).

Motivated by these, in this paper, we present a formalism to study linear and nonlinear optical effects of magnets relying on their spin Hamiltonians. To this end, we systematically derive expressions for the electric polarization represented with spin operators in Hubbard-type systems, by using Schrieffer-Wolff transformation and degenerate perturbation theory~\cite{fazekas1999lecture} with applied electric fields taken into account \cite{Kitamura17,Takasan19,Furuya21}. Combining Green's function approach for magnon excitations (via Holstein-Primakoff transformation) and the obtained expressions for electric polarization, we derive formulae for linear and nonlinear optical conductivities based on the diagrammatic method which was previously used to study nonlinear optical responses of electronic excitations \cite{Parker19}. We apply this method to Heisenberg model and $J_1$-$J_2$ model with broken inversion symmetry. We find that these spin systems support shift current responses due to the superexchange mechanism. Since the present mechanism does not require spin-orbit coupling that is small and usually suppresses the optical responses, such shift current response has a potential to exhibit large nonlinear functionality. One interesting application of such shift current response would be a photodetector that works in far infrared/THz regime.

The rest of this paper is organized as follows. In Sec. II, we present spin Hamiltonian and polarization operator, taking 1D Rice-Mele model with Hubbard interaction as an example. In Sec. II, we study magnon excitations using Holstein-Primakoff transformation and derive its Green's function. In Sec. III, we derive formulae for linear and nonlinear optical conductivities based on the diagrammatic approach. In Sec. IV, we apply our formulae to optical responses in Heisenberg model with alternating coupling constants and $J_1$-$J_2$ model, and demonstrate their shift current responses. In Sec. V, we give a brief discussion.

\section{Polarization in spin systems}
In this section, we derive electric polarization in spin systems. First, we review electric polarization for a Heisenberg model obtained as a low-energy effective theory of Rice Mele model with Hubbard interaction \cite{Katsura09,Tokura14}. Then we present our systematic derivation of polarization operator in spin systems with Schrieffer-Wolff transformation.

\subsection{Rice Mele Hubbard model}
We start from Rice Mele model with Hubbard interaction $U$ (Rice Mele Hubbard model) and derive an effective spin model in the Mott insulator phase. 
This model was previously studied in Ref.~\cite{Katsura09,Tokura14} and the polarization operator has been derived.
The Rice Mele model is a representative 1D model of ferroelectrics~\cite{rice-mele}
which breaks inversion symmetry.
We introduce Hubbard interaction $U$ to this Rice Mele model and derive a spin model that lacks inversion symmetry as its low energy effective theory. The derived spin model is one of the simplest models to study electric polarization and nonlinear optical effects using spin operators.

First we consider the Hamiltonian
\begin{align}
H_0&=\sum_{i,s} \{
[(t +(-1)^i \delta t) c_{i+1,s}^\dagger c_{i,s} +h.c.]
+ (-1)^i m c_{i,s}^\dagger c_{i,s} 
\} \n
&+ U \sum_i  n_{i,\uparrow} n_{i,\downarrow},
\end{align}
where $c_{i,s}$ is the annihilation operator of the electron at site $i$ and spin $s=\uparrow,\downarrow$, 
$n_{i,s}=c_{i,s}^\dagger c_{i,s}$ is the density operator of electrons. $t$ is the overall hopping strength, $\delta t$ is the hopping alternation, $m$ is the staggered potential, and $U$ is the repulsive Hubbard interaction, respectively.
When the Hubbard interaction is sufficiently large, the ground state is in the Mott insulator phase, where each site is occupied with a single electron on average.
While the charge excitation costs large energy ($\simeq U$),
the spin excitations are allowed at low energy. 
We can obtain the effective spin Hamiltonian that describes the low energy spin excitations by perturbation theory with respect to the hopping terms.
Starting from the unperturbed state where each site is occupied with a single electron, we consider the perturbative process where an electron at the site $i$ hops to the site $i+1$ and then hops back to the site $i$. 
This process gives the Heisenberg interaction term,
\begin{align}
2\frac{[t +(-1)^i \delta t]^2}{U - (-1)^i 2m} \bm{S}_i\cdot\bm{S}_{i+1},
\end{align}
where the hopping amplitude is $t +(-1)^i \delta t$ and the intermediate state costs the energy of $U - (-1)^i 2m$.  The Heisenberg coupling $2\bm{S}_i\cdot \bm{S}_{i+1}$ arises as this process is allowed only when the spins at the site $i$ and $i+1$ are antiparallel.
Collecting such contributions from the second order perturbation in $t$, we obtain Heisenberg Hamiltonian
\begin{align}
H&=\sum_i J_i \bm{S}_i\cdot\bm{S}_{i+1},
\label{eq: AF Heisenberg}
\end{align}
with alternating coupling constants
\begin{align}
J_{2i}&= 2(t +\delta t)^2 \left(\frac{1}{U - 2m} + \frac{1}{U + 2m} \right), \\
J_{2i+1}&= 2(t - \delta t)^2 \left(\frac{1}{U - 2m} + \frac{1}{U + 2m} \right).
\end{align}

Next we consider an effective spin Hamiltonian in the presence of an external electric field $E$.
We introduce electrostatic potential to the original Hamiltonian as
\begin{align}
H_E&=H_0 + \sum_{i,s} i \tilde E  c_{i,s}^\dagger c_{i,s},
\end{align}
where we defined the potential difference $\tilde E=eEa$ between the neighboring sites with the electric charge $e$ and the lattice constant $a$.
(Hereafter, we set $e=1$ and $\hbar=1$ for simplicity. The charge of an electron is given by $-e$.)
The change of the onsite potential modifies the spin interaction term that arises from the perturbation process, where an electron hops from the site $i$ to $i+1$ and then hops back to $i$, as
\begin{align}
2\frac{[t +(-1)^i \delta t]^2}{U - (-1)^i 2m + \tilde{E}} \bm{S}_i\cdot\bm{S}_{i+1}.
\end{align}
The total spin Hamiltonian is given by
\begin{align}
\tilde H&=\sum_i \tilde J_i \bm{S}_i\cdot\bm{S}_{i+1},
\end{align}
with
\begin{subequations}
\begin{align}
\tilde J_{2i}&= 2(t +\delta t)^2 \left(\frac{1}{U - 2m + \tilde E} + \frac{1}{U + 2m - \tilde E} \right), \\
\tilde J_{2i+1}&= 2(t -\delta t)^2 \left(\frac{1}{U - 2m - \tilde E} + \frac{1}{U + 2m + \tilde E} \right).
\end{align}
\label{eq: Ji}
\end{subequations}
Taylor expansion with respect to $\tilde E$ gives
\begin{align}
\tilde H&=\sum_i (J_i \bm{S}_i\cdot\bm{S}_{i+1} + E \Pi_i \bm{S}_i\cdot\bm{S}_{i+1}) +O(E^2),\label{eq:Heisenberg-withP}
\end{align}
with
\begin{subequations}
\begin{align}
\Pi_{2i} &= 2a(t +\delta t)^2 \left(-\frac{1}{(U - 2m)^2} + \frac{1}{(U + 2m)^2} \right) , \\
\Pi_{2i+1} &= 2a(t -\delta t)^2 \left(\frac{1}{(U - 2m)^2} - \frac{1}{(U + 2m)^2} \right) .
\end{align}
\label{eq: pi}
\end{subequations}
This equation indicates that the $E$ linear coupling term vanishes when $m=0$, and becomes asymmetric depending on the parity of $i$ as $\Pi_{2i}=-\Pi_{2i+1}$ when $\delta t=0$.

Equation (\ref{eq:Heisenberg-withP}) implies that the electric polarization
of the system is given by
\begin{align}
    P=-\sum_{i}\Pi_{i}\bm{S}_{i}\cdot\bm{S}_{i+1}.
\end{align}
Indeed, by evaluating the expectation value of the electric polarization
$-\sum_{i,s}ia\langle c_{i,s}^{\dagger}c_{i,s}\rangle$ using the
perturbative correction to the electronic eigenvectors, we can confirm that it
coincides with $-\sum_{i}\Pi_{i}\langle\bm{S}_{i}\cdot\bm{S}_{i+1}\rangle$ evaluated in the spin system. In the following, we extend
this relation to general Hubbard-type systems (and higher orders of $E$) using Schrieffer-Wolff
transformation in the presence of applied electric fields.

\subsection{Derivation of polarization operator in spin systems \label{sec: derivation}}

For general Hubbard type systems, we can derive the electric polarization in terms of spin operators in a similar way. 
We employ Schrieffer-Wolff transformation and degenerate perturbation theory \cite{fazekas1999lecture} to deduce the polarization operators in spin operators from the underlying electronic Hamiltonians in the following. 

We show below that the electric polarization of generic
Hubbard-type systems, 
\begin{equation}
\bm{P}_\text{el}=-\sum_{i,s}\bm{R}_{i}c_{i,s}^{\dagger}c_{i,s},
\end{equation}
is expressed in the low-energy spin
description by a simple formula
\begin{equation}
\bm{P}_{\text{spin}}=\hat{P}_0\hat{U}^{\dagger}\bm{P}_\text{el}\hat{U}\hat{P}_0\sim -\dfrac{\partial H_{\text{spin}}}{\partial\bm{E}}.
\label{eq:polarization-formula}
\end{equation}
Here, $\hat{U}$ is the unitary transformation from Hubbard to spin Hamiltonian
in the presence of the (dc) electric field~\cite{Kitamura17,Takasan19,Furuya21}, and $\hat{P}_0$ denotes the projection operator to the spin Hilbert space. Specifically, $\hat{U}$ is defined to satisfy
\begin{equation}
[H_{\text{eff}}, D] = 0,
\end{equation}
with
\begin{align}
H_{\text{eff}} &= \hat{U}^{\dagger}(H_{\text{el}}-\bm{E}\cdot\bm{P}_\text{el})\hat{U}, \\
D&=\sum_{i}n_{i,\uparrow}n_{i,\downarrow}.
\end{align}
Namely, $H_{\text{eff}}$ is the block-diagonalized Hamiltonian that preserves the number of doubly-occupied sites $D$, and its sector with no mobile charge (doubly-occupied or empty sites) corresponds to the spin Hamiltonian $H_\text{spin}=\hat{P}_0H_\text{eff}\hat{P}_0$.

Since the low-energy eigenstate of the Hubbard model, $H_{\text{el}}|\alpha_{\text{el}}\rangle=\varepsilon_{\alpha}|\alpha_{\text{el}}\rangle$,
is given as $|\alpha_{\text{el}}\rangle=\hat{U}|\alpha_{\text{spin}}\rangle$ with $H_{\text{spin}}|\alpha_{\text{spin}}\rangle=\varepsilon_{\alpha}|\alpha_{\text{spin}}\rangle$,
we can show that 
\begin{equation}
\langle\alpha_{\text{el}}|\bm{P}_\text{el}|\alpha_{\text{el}}\rangle=\langle\alpha_{\text{spin}}|\hat{U}^{\dagger}\bm{P}_\text{el}\hat{U}|\alpha_{\text{spin}}\rangle.
\end{equation}
Namely, $\bm{P}_{\text{spin}}=\hat{P}_0\hat{U}^{\dagger}\bm{P}_\text{el}\hat{U}\hat{P}_0$ is indeed the effective
operator to describe the polarization of the underlying electronic
system~\cite{Bulaevskii08}. 
This operator is related to the
$\bm E$ derivative of the spin Hamiltonian as 
\begin{align}
\dfrac{\partial H_{\text{spin}}}{\partial\bm{E}} & =\hat{P}_0\dfrac{\partial}{\partial\bm{E}}\left[\hat{U}^{\dagger}(H_{\text{el}}-\bm{E}\cdot\bm{P}_\text{el})\hat{U}\right]\hat{P}_0\\
 & =-\hat{P}_0\hat{U}^{\dagger}\bm{P}_\text{el}\hat{U}\hat{P}_0+\left[H_{\text{spin}},\hat{P}_0\hat{U}^{\dagger}\partial_{\bm{E}}\hat{U}\hat{P}_0\right],\label{eq:polarization-derivation}
\end{align}
where we have used $\hat{U}^{\dagger}\hat{U}=1$, $\partial_{\bm{E}}\hat{U}^{\dagger}=-\hat{U}^{\dagger}(\partial_{\bm{E}}\hat{U})\hat{U}^{\dagger}$.
We can confirm the formula (\ref{eq:polarization-formula}) by showing that the last term is
 small compared with $\partial_{\bm{E}}H_{\text{spin}}$. 

To this end, let us write the Hubbard-type Hamiltonian as $H_{\text{el}}=\lambda H_{\text{hop}}+H_{\text{loc}}$, where $H_{\text{loc}}$ is the local part of Hamiltonian preserving the number of electrons on each site, while $H_{\text{hop}}$ transfers electrons between
different sites. The dummy parameter $\lambda$ bridges the macroscopically-degenerate atomic limit $\lambda=0$ and the Hubbard system of interest $\lambda=1$, and is formally regarded as a small parameter (as the perturbation $H_{\text{hop}}$ is small).
When the application of $H_{\text{hop}}$ on the
ground state manifold of $H_{\text{loc}}$ costs high energy, we can
perform the perturbation expansion with respect to 
$\lambda$ to obtain the low-energy (spin) Hamiltonian. Namely, by considering a formal expansion 
\begin{align}
\hat{U}&=1+\lambda U^{(1)}+\lambda^{2}U^{(2)}+\dots \equiv e^{-i\Lambda},\\ 
\Lambda&=\lambda \Lambda^{(1)}+\lambda^{2}\Lambda^{(2)}+\dots,     
\end{align}
and imposing that each order of $\hat{U}^\dagger (H_\text{el}-\bm{E}\cdot\bm{P}_\text{el})\hat{U}$ is block-diagonal, we can determine $\hat{U}$ order by order. For example, the first order term $U^{(1)}=-i\Lambda^{(1)}$ is determined by the condition
\begin{equation}
\langle\alpha_{\text{loc}}|(\lambda H_{\text{hop}}+[i\lambda \Lambda^{(1)},H_{\text{loc}}-\bm{E}\cdot\bm{P}_\text{el}])|\beta_{\text{loc}}\rangle=0\label{eq:condition-1st-order}
\end{equation}
for the eigenstates $|\alpha_{\text{loc}}\rangle, |\beta_{\text{loc}}\rangle$ 
of $H_{\text{loc}}-\bm{E}\cdot\bm{P}_\text{el}$ that have different eigenenergies 
$\varepsilon_{\alpha}^{(0)}\neq\varepsilon_{\beta}^{(0)}$.
We choose $\Lambda$ that leaves intrasector matrix elements intact, i.e.,  $\langle\alpha|\Lambda|\beta\rangle=0$ for  $\varepsilon_{\alpha}^{(0)}=\varepsilon_{\beta}^{(0)}$. Thus we arrive at
\begin{equation}
\langle\alpha_{\text{loc}}|\hat{U}^{(1)}|\beta_{\text{loc}}\rangle=
\begin{cases}
-\dfrac{\langle\alpha_{\text{loc}}|H_{\text{hop}}|\beta_{\text{loc}}\rangle}{\varepsilon_{\alpha}^{(0)}-\varepsilon_{\beta}^{(0)}} & \varepsilon_{\alpha}^{(0)}\neq\varepsilon_{\beta}^{(0)}\\
0 & \varepsilon_{\alpha}^{(0)}=\varepsilon_{\beta}^{(0)}
\end{cases}.\label{eq:unitary-1st}
\end{equation}
This solution implies that $\hat{U}^{(1)}$ is a block-offdiagonal matrix (in particular, $\hat{P}_0\hat{U}^{(1)}\hat{P}_0=0$). In other words, $\hat{U}^{(1)}$ always involves charge excitations.
Combining this property with the perturbative evaluation of  $\hat{U}^{\dagger}\partial_{\bm{E}}\hat{U}$,
\begin{equation}
\hat{U}^{\dagger}\partial_{\bm{E}}\hat{U}=\lambda\partial_{\bm{E}}\hat{U}^{(1)}+\lambda^{2}(\hat{U}^{(1)\dagger}\partial_{\bm{E}}\hat{U}^{(1)}+\partial_{\bm{E}}\hat{U}^{(2)})+\dots,
\end{equation}
we find that the first term 
vanishes on the Hilbert space of the spin Hamiltonian, $\lambda\hat{P}_0\partial_{\bm{E}}\hat{U}^{(1)}\hat{P}_0=0$. Thus the last
term of Eq.~(\ref{eq:polarization-derivation}) is higher-order than $H_{\text{spin}}$ by
(at least) $\lambda^{2}$, while $H_{\text{spin}}$ and $\partial_{\bm{E}}H_{\text{spin}}$
is usually in the same order in $\lambda$. This leads to Eq.~(\ref{eq:polarization-formula}).

As an example, let us consider a single-orbital Hubbard model with
an arbitrary onsite potential
\begin{subequations}
\begin{align}
H_{\text{hop}} & =\sum_{ijs}t_{ij}c_{i,s}^{\dagger}c_{j,s},\\
H_{\text{loc}} & =\sum_{i}[V_{i}(n_{i,\uparrow}+n_{i,\downarrow})+Un_{i,\uparrow}n_{i,\downarrow}].
\end{align}\label{eq:generic-Hubbard}
\end{subequations}
As detailed in Appendix~\ref{app: derivation}, application of the above formalism leads to the effective spin Hamiltonian as
\begin{align}
H_{\text{spin}} & =\dfrac{1}{2}\sum_{ij}\dfrac{4|t_{ij}|^{2}\bm{S}_{i}\cdot\bm{S}_{j}}{U-V_{ij}-\bm{E}\cdot\bm{R}_{ij}},\label{eq:generic-Heisenberg}
\end{align}
and the effective polarization operator as
\begin{align}
\bm{P}_{\text{spin}}&=-\frac{1}{2}\sum_{ij}\dfrac{4|t_{ij}|^{2}\bm{R}_{ij}(\bm{S}_{i}\cdot\bm{S}_{j})}{(U-V_{ij}-\bm{E}\cdot\bm{R}_{ij})^{2}}\label{eq:polarization-hubbard-0}\\
&=-\dfrac{1}{2}\sum_{ij}\dfrac{8UV_{ij}|t_{ij}|^{2}\bm{R}_{ij}(\bm{S}_{i}\cdot\bm{S}_{j})}{(U^{2}-V_{ij}^{2})^{2}}+O(E^2),\label{eq:polarization-hubbard}
\end{align}
in the second-order perturbation, where $\bm{R}_{ij}=\bm{R}_i-\bm{R}_{j}, V_{ij}=V_i-V_j$.
We can indeed confirm that Eq.~(\ref{eq:polarization-formula})
holds true including higher orders of $E$. We can also see that 
it reproduces the result for $P$ in Rice-Mele-Hubbard model;
Eq.~\eqref{eq:polarization-hubbard} coincides with the result for $\Pi_i$ in Eq.~\eqref{eq: pi} in the limit $E \to 0$. 

We note that the formula (\ref{eq:polarization-formula}) slightly deviates when we consider ac electric
fields. 
However, as we show in Appendix~\ref{app: derivation}, it turns out that the correction term in the low-frequency regime scales as $\sim\omega^{2}/U^{2}$ with $\omega$ being the driving frequency, which is negligible since we are interested in optical excitation of magnons.

The rough estimation of the spin-dependent electric polarization $P$ in the Hubbard-type systems is obtained from Eq.~\eqref{eq:polarization-hubbard} as
\begin{align}
    P \simeq \frac{JV_{ij}}{U^2} a,
\end{align}
where $J$ is the Heisenberg coupling ($J\simeq t_{ij}^2/U$) and $a$ is the lattice constant. Note that the difference in the onsite potential $V_{ij}$ cannot be as large as $U$ with keeping the half-filled condition.
The electric polarization in spin systems is enhanced if $P$ is expressed in the lower power in $1/U$. 
In fact, we can realize
\begin{align}
    P \simeq \frac{J}{U} a,
\end{align}
in the so called superexchange mechanism where neighboring spin sites are bridged by ligand ions. Details are presented in Appendix \ref{app: polarization}.
In the superexchange mechanism, the Heisenberg coupling appears in the fourth order perturbation in hopping between spin and ligand sites. Such situation reduces the power of $1/U$ in $P$ and enhances the electric polarization and an effective coupling to external electric fields.

\section{Magnon excitations}
We consider magnon excitations using Holstein-Primakoff transformation.
While 1D Heisenberg model with alternating coupling supports spin gap, the present treatment with Holstein-Primakoff transformation is justified for higher dimensional systems (such as a stack of the Heisenberg chains).

We consider antiferromagnetic ground state and use 
Holstein-Primakoff transformation of spin operators for spin $S$ states given by
\begin{subequations}
\begin{align}
S_{i}^{+}&= \sqrt {2S} \sqrt{1-\frac {a_{i}^{\dagger }a_{i}}{2S}}a_i, \\
S_{i}^{-}&= \sqrt {2S} a_i^{\dagger }\sqrt{1-\frac {a_i^{\dagger }a_i}{2S}}, \\
S_{i}^{z}&= S-a_i^{\dagger }a_i, 
\end{align}
for even sites ($i=2j$)
and
\begin{align}
S_{i}^{+}&= \sqrt {2S} a_i^{\dagger }\sqrt{1-\frac {a_i^{\dagger }a_i}{2S}}, \\
S_{i}^{-}&= \sqrt {2S} \sqrt{1-\frac {a_i^{\dagger }a_i}{2S}}a_i,  \\
S_{i}^{z}&= -S+a_i^{\dagger }a_i, 
\end{align}
\label{eq: HP transf}
\end{subequations}
for odd sites ($i=2j+1$),
with boson annihilation operator $a_i$ at site $i$.
When $S$ is large and boson density is small, these equations reduce to
\begin{align}
S_{i}^{+}&= 
\begin{cases}
\sqrt {2S} a_i, & (i=2j) \\
\sqrt {2S} a_i^\dagger, & (i=2j+1) \\
\end{cases}\\
S_{i}^{z}&= (-1)^i(S-a_i^{\dagger }a_i).
\end{align}

\subsection{Heisenberg model with alternating coupling constants}
We consider the Hamiltonian
\begin{align}
H &= \sum_i (J_e \bm S_{2i} \cdot \bm S_{2i+1} + J_o \bm S_{2i+1} \cdot \bm S_{2i+2}),
\end{align}
which is obtained from Rice Mele Hubbard model by setting $J_e=J_{2i}$ and $J_o=J_{2i+1}$.
We assume that the ground state is an antiferromagnetic state and apply Holstein-Primakoff transformation to study magnetic excitations.
We obtain
\begin{align}
H &= \sum_i J_e S 
\begin{pmatrix}
a_{2i}^\dagger & a_{2i+1}
\end{pmatrix}
\begin{pmatrix}
1 & 1 \\
1 & 1
\end{pmatrix}
\begin{pmatrix}
a_{2i} \\ a_{2i+1}^\dagger 
\end{pmatrix}
\n
&+
\sum_i J_o S 
\begin{pmatrix}
a_{2i+2}^\dagger & a_{2i+1}
\end{pmatrix}
\begin{pmatrix}
1 & 1 \\
1 & 1
\end{pmatrix}
\begin{pmatrix}
a_{2i+2} \\ a_{2i+1}^\dagger 
\end{pmatrix}.
\end{align}
By performing a Fourier transformation, we obtain 
\begin{align}
H&=\sum_q 
\begin{pmatrix}
a_{A,q}^\dagger & a_{B,-q}
\end{pmatrix}
H_q
\begin{pmatrix}
a_{A,q} \\ a_{B,-q}^\dagger 
\end{pmatrix},
\\
H_q&= 2S 
\begin{pmatrix}
J & J \cos qa - i \delta J \sin qa \\
J \cos qa + i \delta J \sin qa & J
\end{pmatrix}
\end{align}
where $a_{A,q}$ and $a_{B,q}$ are annihilation operators of magnons at the even and odd sites with the momentum $q$, and 
$J=(J_e+J_o)/2$ and $\delta J=(-J_e+J_o)/2$.
The Green's function for the magnon excitation is given by
\begin{align}
G(i\omega, q) &= (i\omega \sigma_z - H_q)^{-1}
\end{align}
with Matsubara frequency $i \omega$ and the momentum $q$. Here, $\sigma_z$ appears from the fact that the basis of the two by two Hamiltonian is spanned by an annihilation operator and a creation operator of bosons and incorporates the right signs of the two modes.
Since the poles of the Green's function are given by the eigenvalues of $\sigma_z H_q$, the excitation energy of magnons are
\begin{align}
E&=2SJ \sqrt{1-\cos^2 qa -\frac{\delta J^2}{J^2}\sin^2 qa} \n
&= 2SJ \sqrt{1-\frac{\delta J^2}{J^2}} |\sin qa|.
\end{align}

In the same two by two representation, the modification of the spin Hamiltonian $H \to H - E P$ in the presence of the electric field $E$ defines the polarization operator $P$ as
\begin{align}
P&=
-\sum_q 
\begin{pmatrix}
a_{A,q}^\dagger & a_{B,-q}
\end{pmatrix}
\Pi_q
\begin{pmatrix}
a_{A,q} \\ a_{B,-q}^\dagger 
\end{pmatrix},
\\
\Pi_q&= 2S
\begin{pmatrix}
\Pi & \Pi \cos qa - i \delta \Pi \sin qa \\
\Pi \cos qa + i \delta \Pi \sin qa & \Pi
\end{pmatrix},
\end{align}
which is obtained from Rice Mele Hubbard model by setting 
$\Pi=(\Pi_{2i}+\Pi_{2i+1})/2$ and $\delta \Pi=(-\Pi_{2i}+\Pi_{2i+1})/2$.

\subsection{General magnon Hamiltonian}
Finally, we consider a general magnon Hamiltonian in a bilinear form of magnon operators which is written as
\begin{align}
H&= \sum_q \Psi_q^\dagger H_q \Psi_q,
\label{eq: H general}
\end{align}
with a $2n$ by $2n$ matrix $H_q$ and a $2n$ dimensional vector of bosonic operators $\Psi_q$.
In this general case, the Green's function of magnons is given by
\begin{align}
G(i\omega, q) = (i\omega B - H_q)^{-1} = (i\omega  - B H_q)^{-1} B.
\end{align}
Here $B$ is the diagonal matrix with entries $\eta_i$,
\begin{align}
B\equiv \t{diag}(\eta_i),
\end{align}
where $\eta_i=1$ if the $(\Psi_q)_i$ is an annihilation operator, and
$\eta_i=-1$ if the $(\Psi_q)_i$ is a creation operator. Namely, $B$ is related to the commutation relation as $[(\Psi_{q})_{i},(\Psi_{q}^{\dagger})_{j}]=(B)_{ij}$.
The energy dispersion of magnon excitations corresponds to the poles of $G(i\omega, q)$ and is given by the eigenvalues of $B H_q$. Note that $B^2=I$ where $I$ is the identity matrix.
Let us assume that this matrix can be diagonalized as
\begin{align}
BH_q &= V E_q V^{-1},
\end{align}
where $E_q=\t{diag}(\epsilon_{i})$ with the eigenvalues $\epsilon_{i}$.
The eigenvalues $\epsilon_{i}$ appear as pairs of a positive mode and a negative mode, and the magnon dispersion is determined by the positive eigenvalues of $BH_q$.

Let us remark the relation between the diagonalization using $V$
and that using Bogoliubov transformation. We find that the diagonalized form
of the Hamiltonian reads 
\begin{equation}
H=\sum_{q}(\Psi_{q}^{\dagger}BVB)BE_{q}(V^{-1}\Psi_{q})
\end{equation}
and the transformed operators satisfy the commutation relation
\begin{equation}
[(V^{-1}\Psi_{q})_{i},(\Psi_{q}^{\dagger}BVB)_{j}]=(B)_{ij}.
\end{equation}
Therefore the diagonalized Hamiltonian coincides with the Bogoliubov-transformed
one that satisfies canonical commutation relation, if $(V^{-1}\Psi_{q})^{\dagger}=\Psi_{q}^{\dagger}BVB$ holds.
Namely, the Bogoliubov-transformed result is recovered by further
imposing $V^{\dagger}BV=B$ (while it is not necessary for the following calculations). 
When this condition is met, the diagonal entries of $BE_q$ have physical meaning as the excitation energies.

\section{Optical responses \label{sec: conductivities}}
We study linear optical conductivity and shift current response using the Green's function formalism for magnon excitations.
In particular, we derive expressions for optical conductivities in terms of matrix elements of the polarization operator using the general magnon Hamiltonian.

\subsection{Linear optical conductivity}
We consider the linear optical conductivity $\sigma^{(1)}(\omega)$ which characterizes the current response
\begin{align}
J(\omega) &= \sigma^{(1)}(\omega) E(\omega),
\end{align}
where $J(\omega)$ and $E(\omega)$ are Fourier components of the current and the external electric field, respectively.
Our formalism of magnon excitations with $E$ naturally includes the polarization operator ($H'/E$) in the Hamiltonian.
Since the current $J$ is given by the time derivative of the polarization $P$, we have the relationship $J(\omega)= -i \omega P(\omega)$.
Thus the optical conductivity is obtained from the dielectric function $\epsilon(\omega)$ as
\begin{align}
\sigma^{(1)}(\omega)=-i\omega \epsilon(\omega)
\end{align}
where $\epsilon(\omega)$ satisfies $P(\omega)=\epsilon(\omega) E(\omega)$.
The dielectric function $\epsilon(\omega)$ is given by the two point correlation function of polarization as
\begin{align}
\epsilon(i\Omega)
&=
\int \frac{dq}{2\pi} \int \frac{d\omega}{2\pi} \t{tr}[\Pi_q G(i\omega+i\Omega, q) \Pi_q G(i\omega, q)],
\end{align}
where $q$ is the momentum of the magnon and the $q$ integral is performed over the Brillouin zone.
Here, $\Pi_q$ is a $2n$ by $2n$ matrix that defines the polarization operator as 
\begin{align}
    P&=-\sum_q \Psi_q^\dagger \Pi_q \Psi_q.
\end{align}
(We note that there is an overall extra minus sign in the above expression for the two-point correlation function when compared to the fermionic case, which is canceled with the minus sign in $P=-\sum_q \Pi_q \Psi_q^\dagger \Psi_q$.)

Now we derive an explicit expression for the linear conductivity in terms of the matrix elements of the general magnon Hamiltonian Eq.~\eqref{eq: H general}.
The expression for the linear susceptibility can be rewritten as
\begin{align}
\epsilon(i\Omega)
&=
\int \frac{dq}{2\pi} \int \frac{d\omega}{2\pi} \t{tr}[\Pi_q G(i\omega+i\Omega, q) \Pi_q G(i\omega, q)] \n
&= \int \frac{dq}{2\pi} \int \frac{d\omega}{2\pi}
\t{tr}[\Pi_q V (i\omega+i\Omega -E_q)^{-1} V^{-1} B  \n
&\hspace{7em} \times \Pi_q V (i\omega -E_q)^{-1} V^{-1}B] \n
&= \int \frac{dq}{2\pi} \int \frac{d\omega}{2\pi} 
\sum_{ab} \frac{\widetilde \Pi_{ab} \widetilde \Pi_{ba}}{(i\omega+i\Omega-\epsilon_b)(i\omega-\epsilon_a)} \n
&=\int \frac{dq}{2\pi} \sum_{ab} \widetilde \Pi_{ab} \widetilde \Pi_{ba} 
\frac{f_{ab}}{i\Omega - \epsilon_{ba}}.
\end{align}
The matrix $\widetilde \Pi$ is defined by
\begin{align}
\widetilde \Pi=V^{-1} B \Pi_q V,
\end{align}
and $\widetilde \Pi_{ab}$ is its matrix element.
$f_{ab}=f_a-f_b$ is a factor assuring that the positive energy mode is excited with $f_a \equiv \theta(-\epsilon_a)$,
and $\epsilon_{ab}=\epsilon_a-\epsilon_b$.
By performing analytic continuation of Matsubara frequency $i\Omega \to \omega + i\gamma$,
we obtain the expression for the linear conductivity of magnons as
\begin{align}
\sigma^{(1)}(\omega) &=
-i\omega \int \frac{dq}{2\pi} \sum_{ab} \widetilde \Pi_{ab} \widetilde \Pi_{ba} 
\frac{f_{ab}}{\omega - \epsilon_{ba}+i\gamma}.
\label{eq: sigma1}
\end{align}
We note that the matrix $\widetilde \Pi$ is not Hermitian and $\widetilde \Pi_{ab} \neq \widetilde \Pi_{ba}^*$ generally.

\subsection{Nonlinear optical conductivity}
The second order nonlinear conductivity is defined by the current response
\begin{align}
J(\omega_1+\omega_2) &= 
\sigma^{(2)}(\omega_1+\omega_2; \omega_1, \omega_2) E(\omega_1) E(\omega_2),
\end{align}
where the external electric fields of the frequencies $\omega_1$ and $\omega_2$ yields the current of the sum frequency $\omega_1+\omega_2$.
Similarly to the case of the linear response,
nonlinear conductivity is related to the nonlinear susceptibility via
\begin{align}
\sigma^{(2)}(\omega_1+\omega_2; \omega_1, \omega_2) &=
-i(\omega_1+\omega_2)\chi(\omega_1+\omega_2; \omega_1, \omega_2),
\end{align}
where the nonlinear susceptibility characterizes the nonlinear response of polarization as
\begin{align}
P(\omega_1+\omega_2) &= 
\chi(\omega_1+\omega_2; \omega_1, \omega_2) E(\omega_1) E(\omega_2).
\end{align}

The shift current is the generation of dc current flow proportional to the intensity of the light field in noncentrosymmetric crystals.
Such current response is described by the nonlinear conductivity in the dc response limit,
\begin{align}
\sigma^{(2)}(\omega) \equiv \lim_{\delta \omega \to 0} \sigma^{(2)}(\delta \omega; \omega + \delta \omega, -\omega).
\end{align}
In terms of nonlinear susceptibility, we can express the shift current response as
\begin{align}
\sigma^{(2)}(\omega) &= \lim_{\delta \omega \to 0}
(-i\delta \omega) \chi(\delta \omega; \omega + \delta \omega, -\omega).
\end{align}
The nonlinear susceptibility $\chi$ is contributed by the two point correlation function of $\Pi_q$ and $\partial_q \Pi_q$ 
and the three point correlation function of $\Pi_q$.
The above expression for $\sigma^{(2)}$ indicates that the part of $\chi$ that is proportional to $1/\delta\omega$ makes a contribution. Such $1/\delta \omega$ term appears in the three point correlation function $\chi^{(3)}$ of $\Pi_q$. (The two point correlation part does not include a singular part with respect to $\delta \omega$ and vanishes after taking the limit.)
Therefore, the shift current response is given by
\begin{align}
\sigma^{(2)}(\omega) &= \lim_{\delta \omega \to 0}
(-i\delta \omega) \chi^{(3)}(\delta \omega; \omega + \delta \omega, -\omega),
\end{align}
where the three point correlation function $\chi^{(3)}$ is written as
\begin{align}
&\chi^{(3)}(i\Omega_1+i\Omega_2; i\Omega_1, i\Omega_2) \n
&=
\int \frac{dq}{2\pi} \int \frac{d\omega}{2\pi} 
\n & \quad 
\t{tr}\Big[ \Pi_q G(i\omega+i\Omega_1+i\Omega_2, q) \Pi_q G(i\omega+i\Omega_1, q) \Pi_q G(i\omega, q) \n
& 
\qquad +\Pi_q G(i\omega+i\Omega_1+i\Omega_2, q) \Pi_q G(i\omega+i\Omega_2, q) \Pi_q G(i\omega, q)
\Big].
\end{align}
We note that we perform analytic continuation of Matsubara frequencies as 
\begin{align}
i\Omega_1 &\to \omega + \delta \omega+  i\gamma, & 
i\Omega_2 &\to - \omega +  i\gamma, 
\end{align}
where $\gamma$ corresponds to the energy broadening and it enters with plus signs from causality.

Now that we sketched the overview of the derivation of $\sigma^{(2)}(\omega)$, we derive the explicit expression for the nonlinear conductivity $\sigma^{(2)}(\omega)$ in terms of the matrix elements for the general magnon Hamiltonian in Eq.~\eqref{eq: H general}.
The nonlinear susceptibility $\chi^{(3)}$ can be expressed as
\begin{align}
&\chi^{(3)}(i\Omega_1+i\Omega_2; i\Omega_1, i\Omega_2) \n
&=
\int \frac{dq}{2\pi} \int \frac{d\omega}{2\pi} 
\sum_{abc}\widetilde \Pi_{ac} \widetilde \Pi_{cb} \widetilde \Pi_{ba} \n
&\times 
\Big\{
\frac{1}{
(i\omega+i\Omega_1+i\Omega_2 -\epsilon_c)(i\omega+i\Omega_1-\epsilon_b)(i\omega-\epsilon_a)
}
\n
&+
\frac{1}{
(i\omega+i\Omega_1+i\Omega_2 -\epsilon_c)(i\omega+i\Omega_2-\epsilon_b)(i\omega-\epsilon_a)
}
\Big\} \n
&= \int \frac{dq}{2\pi} \sum_{abc}\widetilde \Pi_{ac} \widetilde \Pi_{cb} \widetilde \Pi_{ba} \n
& 
\times \Big\{
\frac{f_a}{(i\Omega_1-\epsilon_{ba})(i\Omega_1+i\Omega_2-\epsilon_{ca})} 
- \frac{f_b}{(i\Omega_1-\epsilon_{ba})(i\Omega_2-\epsilon_{cb})} \n
&+ \frac{f_c}{(i\Omega_1+i\Omega_2-\epsilon_{ca})(i\Omega_2-\epsilon_{cb})}
\Big\} 
 + (i\Omega_1 \leftrightarrow i\Omega_2).
\end{align}
Among the terms in the above expressions, we are interested in the pieces proportional to $1/\delta \omega$ as we analytically continue as
\begin{align}
i\Omega_1 &\to \omega +\delta \omega + i\gamma, \n i\Omega_2 &\to -\omega + i\gamma, \label{eq: analytic continuation} \\
i\Omega_1 + i\Omega_2 &\to \delta \omega + 2i\gamma. \nonumber
\end{align} 
The factor $1/\delta \omega$ arises from $1/(i\Omega_1+i\Omega_2-\epsilon_{ca})$ by setting $c=a$ which leads to $1/(\delta \omega + 2i \gamma)$. 
(Here we assume $\gamma \ll \delta \omega$. 
We further discuss the effects of dissipation in Sec. \ref{sec: discussion}.)
Collecting those terms, we obtain
\begin{align}
&\chi^{(3)}(i\Omega_1+i\Omega_2; i\Omega_1, i\Omega_2) \n
&=\int \frac{dq}{2\pi} \sum_{ab}
\frac{\widetilde \Pi_{aa} \widetilde \Pi_{ab} \widetilde \Pi_{ba}}{(i\Omega_1 + i\Omega_2)}
\Big\{
\left( \frac{f_{ab}}{i\Omega_1 - \epsilon_{ba}} + \frac{f_{ab}}{i\Omega_2 - \epsilon_{ab}} \right) \n
&+\left( \frac{f_{ab}}{i\Omega_2 - \epsilon_{ba}} + \frac{f_{ab}}{i\Omega_1 - \epsilon_{ab}} \right)
\Big\} + O(1),
\label{eq: chi (3)}
\end{align}
which yields
\begin{align}
&\lim_{\delta\omega\to 0} \delta\omega \chi^{(3)}(\delta\omega+2i\gamma; \omega +\delta \omega + i\gamma, -\omega + i\gamma) \n
&= \int \frac{dq}{2\pi} \sum_{ab}
\widetilde \Pi_{ab} \widetilde \Pi_{ba} (\widetilde \Pi_{aa} - \widetilde \Pi_{bb}) \n
&\qquad\qquad \times
\left( \frac{f_{ab}}{ \omega - \epsilon_{ba} + i\gamma} + \frac{f_{ab}}{-\omega - \epsilon_{ab} + i \gamma} \right).
\end{align}
Focusing on the optical excitation of magnons, we obtain
\begin{align}
&\sigma^{(2)}(\omega) \n
&=
-2\pi \int \frac{dq}{2\pi} \sum_{ab} \t{Re}[\widetilde \Pi_{ab} \widetilde \Pi_{ba} (\widetilde \Pi_{aa} - \widetilde \Pi_{bb})] f_{ab} \delta(\omega-\epsilon_{ba}).
\label{eq: sigma2}
\end{align}
In the above expression, the factor $\widetilde \Pi_{aa} - \widetilde \Pi_{bb}$ corresponds to the polarization difference for the two magnon modes labeled by $a$ and $b$, and can be regarded as a counterpart of shift vector for the electronic shift current which is a geometric quantity involving Berry connection of Bloch electrons.

\begin{figure*}
\begin{center}
\includegraphics[width=0.8\linewidth]{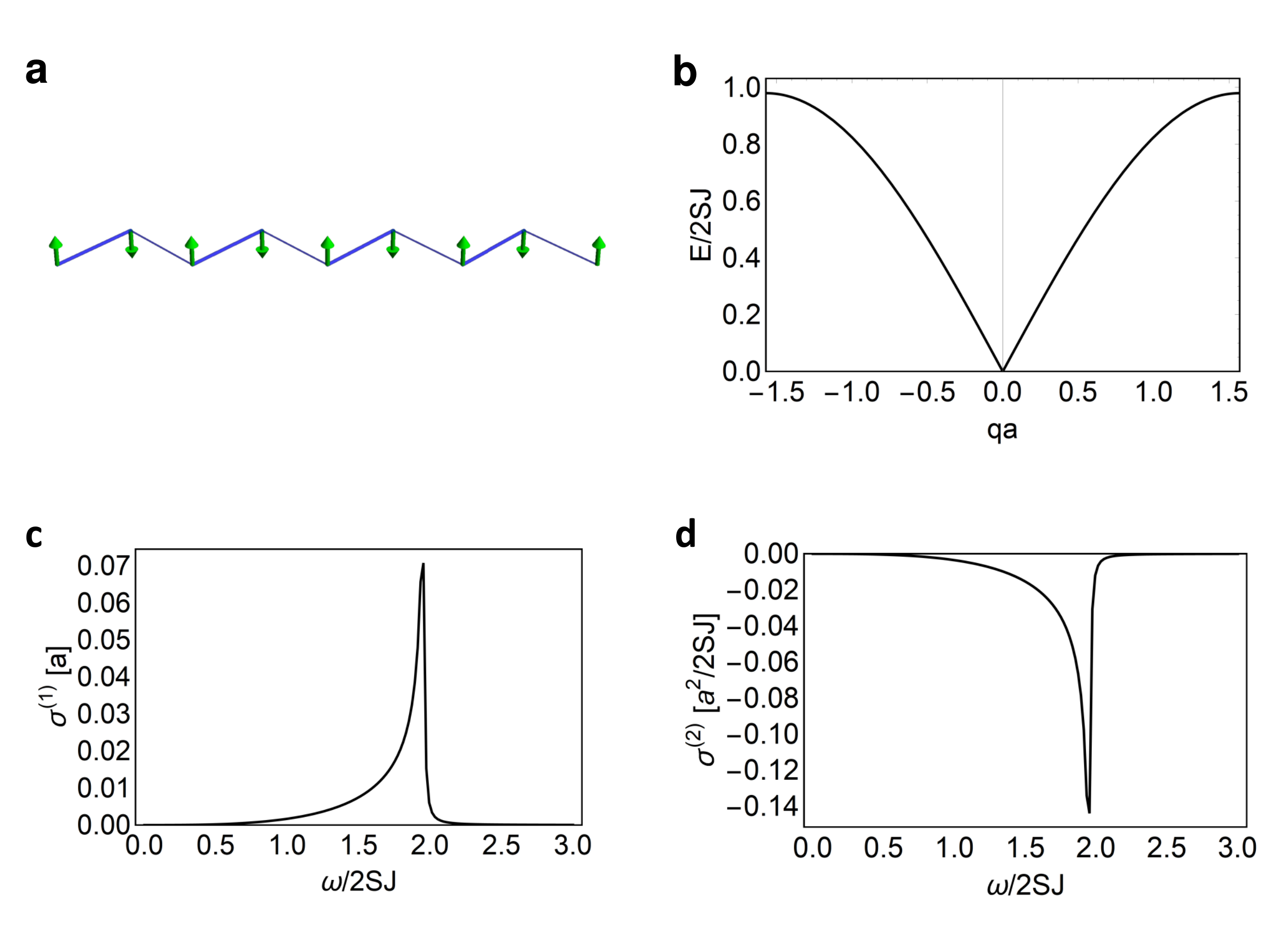}
\caption{\label{fig: sigma 12}
Optical responses of Heisenberg model with broken inversion symmetry.
(a) A schematic picture of the antiferromagnetic state in inversion broken Heisenberg model.
(b) Magnon dispersion.
(c) Linear optical conductivity $\sigma^{(1)}(\omega)$. The peak at $\omega/2SJ=1.9$ corresponds to the optical excitation of magnons at the zone boundary $qa=\pi/2$.
(d) Nonlinear conductivity $\sigma^{(2)}(\omega)$ characterizing shift current response. 
We used parameters $2S(J, \delta J)=(1,0.2), 2S(\Pi, \delta \Pi) =(1,0.1)a, g=0.01$.
}
\end{center}
\end{figure*}

\section{Applications}
In this section, we apply our formulation of linear and nonlinear conductivities to various spins systems that break inversion symmetry and show that shift current emerges by magnon excitations in those systems.

\subsection{Heisenberg model with alternating coupling constants}
First we consider optical responses in the Heisenberg model with alternating coupling constants.
In this case, we can derive analytical expressions for linear and nonlinear conductivities from Eq.~\eqref{eq: sigma1} and Eq.~\eqref{eq: sigma2} as 
\begin{align}
    \sigma^{(1)}(\omega)&=\frac{ (\omega/2S) ^3 (J\delta \Pi -\Pi\delta J)^2}{4a(J^2-\delta J^2)^2 \sqrt{4( J^2- \delta J^2)-(\omega/2S) ^2}},\\
    \sigma^{(2)}(\omega)&=-\frac{  (\omega/2S) ^3 (J\delta \Pi- \Pi\delta J)^2 (J \Pi-\delta J \delta \Pi)}{2a(J^2-\delta J^2)^3 \sqrt{4 (J^2- \delta J^2)-(\omega/2S) ^2}}.
\end{align}
Figure \ref{fig: sigma 12} shows the magnon spectrum, linear conductivity $\sigma^{(1)}(\omega)$, and nonlinear conductivity $\sigma^{(2)}(\omega)$.
Magnon excitation shows a linear dispersion around $q=0$ and band bending at $q=\pi/2a$ (Fig.~\ref{fig: sigma 12}(b)). The linear conductivity shows a peak structure at $\omega/2S=2 J\sqrt{1-\delta J^2/J^2}$ (Fig.~\ref{fig: sigma 12}(c)),
which is associated with two magnon excitations around the zone boundary ($q=\pi/2a$) due to the large density of states of magnons at $q=\pi/2a$.
We find that the nonlinear conductivity $\sigma^{(2)}$ is nonzero as a consequence of inversion breaking encoded in the polarization operators (Fig.~\ref{fig: sigma 12}(d)). $\sigma^{(2)}$ also shows a peak structure at $\omega/2S=2 J\sqrt{1-\delta J^2/J^2}$ which is again associated with two magnon
excitations around the zone boundary. This nonlinear response can be interpreted as follows.
(i) Light irradiation excites magnons due to the coupling term proportional to $E$. (ii) Magnons accompany nonzero polarization due to inversion symmetry breaking (which are so called electromagnons). (iii) Constant light irradiation induces increase of polarization from magnon excitations which drives dc current flow.
We can see that inversion symmetry breaking is necessary for nonvanishing nonlinear conductivity. For example, when the site center inversion symmetry is present, we have $\delta t=0$ indicating $\delta J=0, \Pi=0$, for which we can verify $\sigma^{(2)}(\omega)$ vanish. Similarly, the bond center inversion symmetry requires $m=0$, which constrains $\Pi=\delta \Pi=0$ and naturally leads to $\sigma^{(2)}(\omega)=0$.

\begin{figure*}
\begin{center}
\includegraphics[width=0.8\linewidth]{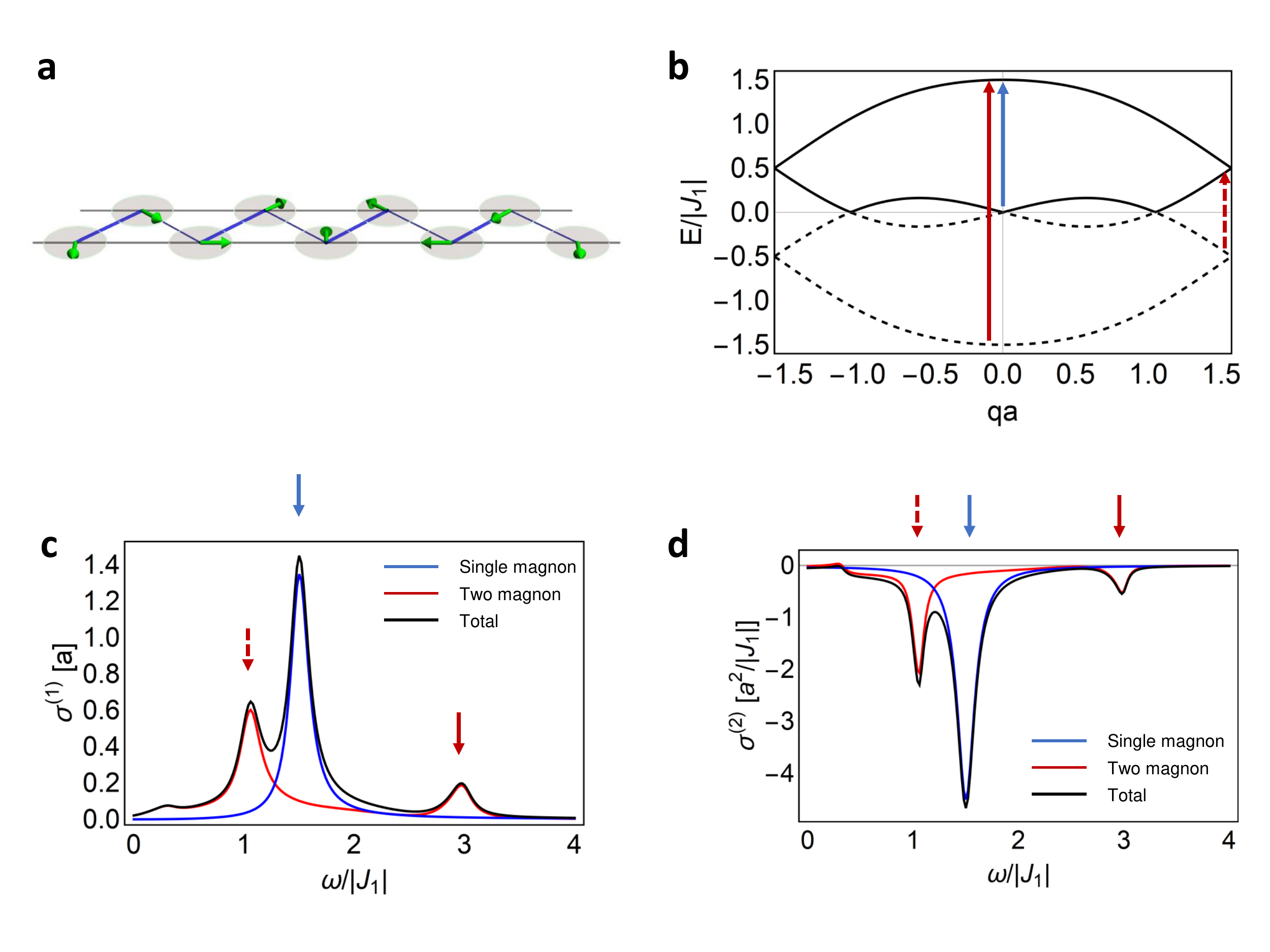}
\caption{\label{fig: J1J2}
Optical responses of $J_1$-$J_2$ spin chains with broken inversion symmetry.
(a) A schematic picture of the cycloidal spin structure.
(b) Magnon dispersion. The solid (dashed) lines represent positive (negative) energy modes. Three arrows represent dominant optical excitations (blue for single magnon resonance, and red for two magnon resonances).
(c) Linear optical conductivity $\sigma^{(1)}(\omega)$. Blue, red and black curves represent the single magnon contribution, the two magnon contribution, and the total conductivity, respectively. Three peaks correspond to the three dominant optical excitations where the joint density of states is large.
(d) Nonlinear conductivity $\sigma^{(2)}(\omega)$ characterizing shift current response. Blue, red and black curves represent the single magnon contribution, the two magnon contribution, and the total conductivity, respectively.
We used parameters $(J_1, J_2)=(-1, 0.5), (\Pi_1,\delta \Pi_1)=(-1, 0.2)a, S=1, g=0.1$.
}
\end{center}
\end{figure*}

\subsection{$J_1$-$J_2$ spin chains}
We consider $J_1$-$J_2$ spin chain which is described by the Hamiltonian,
\begin{align}
H&=J_1 \sum_i \bm{S}_i \cdot \bm{S}_{i+1} 
+ J_2 \sum_i \bm{S}_i \cdot \bm{S}_{i+2}.
\label{eq: H J1J2}
\end{align}
For $J_1<0$ and $J_2>0$, the ground state exhibits the cycloidal spin structure of
\begin{align}
\bm{S}_i = S (\sin Q i a, 0, \cos Q i a)
\label{eq: cycloidal}
\end{align}
with the momentum $Q$ given by $\cos Qa = -J_1/4J_2$  \cite{Yoshimori59}.
We adopt the polarization operator 
\begin{align}
P&= -\sum_i \Pi_i \bm{S}_i \cdot \bm{S}_{i+1} \n
&=- \sum_i \{ \Pi_1 + (-1)^i \delta \Pi_1 \} \bm{S}_i \cdot \bm{S}_{i+1}, \label{eq:P-J1J2}
\end{align}
which consists of uniform Heisenberg term $\Pi_1$ and coupling alternation $\delta \Pi_1$ between the nearest spins.
This form of polarization operator arises from exchange striction mechanism in general.
Indeed we present an explicit derivation of this form of the polarization operator in Appendix \ref{app: polarization}, where we consider effects of ligand ions that bridge spin sites and perform perturbation theory incorporating exchange processes involving such nonmagnetic bridging sites.
We note that the direction of $P$ is model dependent. In the following, we choose the directions of the electric field of the incident light and the induced dc current to be the same as that of $P$.

We study the magnetic excitations and the associated current responses by using the magnon representation for the Hamiltonian and the polarization operator. 
If we consider the magnon excitations from the cycloidal spin structure in Eq.~\eqref{eq: cycloidal},
the operator $\mathcal{O}$ defined by
\begin{align}
\mathcal{O}&=
\sum_i \{c_1 + \delta c_1 (-1)^i\} \bm{S}_i \cdot \bm{S}_{i+1} 
+ c_2 \sum_i \bm{S}_i \cdot \bm{S}_{i+2}
\end{align}
can be represented with magnon operators as  \cite{Miyahara12,Hasegawa10}
\begin{align}
\mathcal{O}&=
\sum_q
\Psi_q^\dagger
\begin{pmatrix}
A_q &  B_q & C_{-q} &  D_{-q} \\
B_q &  A_q & D_{-q} &  C_{-q} \\
C_{q} &  D_{q} & A_q &  B_q  \\
D_{q} &  C_{q} & B_q &  A_q  \\
\end{pmatrix}
\Psi_q + u_0\Psi_{q=0}, 
\label{eq: a rep for O}
\end{align}
with
\begin{align}
\Psi_q &=
\begin{pmatrix}
a_{qA} \\  a_{-qA}^\dagger \\ a_{qB} \\ a_{-qB}^\dagger
\end{pmatrix},
\end{align}
where $a_{qA/B}$ is the annihilation operator of magnon with the momentum $q$ on the sublattice A/B (sublattice A: odd sites, B: even sites), and the coefficients are given by
\begin{align}
A_q &=-c_1 S\cos Qa + c_2 S\{\cos^2 Qa \cos 2q - \cos 2Qa \}, \\
B_q &= -c_2 S\sin^2Qa \cos 2qa, \\
C_q &= S\cos^2\frac {Qa} 2 (c_1 \cos qa +i\delta c_1 \sin qa), \\
D_q &= -S\sin^2\frac {Qa} 2 (c_1 \cos qa +i\delta c_1 \sin qa),
\end{align}
and
\begin{align}
u_0 &= 2S\sqrt{\frac{S}{2}}\delta c_1\sin{Qa} (1,1,-1,-1).
\label{eq: u0}
\end{align}
(For derivation, see Appendix~\ref{app: j1j2}).
Using this formula, we can obtain magnon representation for $H$ by setting $(c_1, \delta c_1, c_2) \to (J_1, 0, J_2)$,
which gives only a bilinear form in magnon operators ($u_0=0$).
We obtain the energy dispersion of magnon excitations by diagonalizing $H$ thus obtained.
Similarly, magnon representation for the electric polarization is obtained by replacing $(c_1, \delta c_1, c_2) \to (\Pi_1, \delta \Pi_1, 0)$ with an overall minus sign,
which reads
\begin{align}
    P=-\Big[\sum_q \Psi_q^\dagger \Pi_q \Psi_q + \pi_0 \Psi_{q=0}\Big].
\end{align}
Specifically, the first term in the right hand side of Eq.~\eqref{eq: a rep for O} gives the bilinear part with $\Pi_q$, and the second term gives the single magnon term with $\pi_0$.
The bilinear part with $\Pi_q$ leads to the two magnon contribution to optical conductivities as we formulated in Sec.~\ref{sec: conductivities}. In contrast, the single magnon term with $\pi_0$ gives rise to the single magnon contribution to the optical conductivities which requires a separate treatment as we detail in Appendix~\ref{app: j1j2}.

We show the magnon dispersion and the linear and nonlinear conductivities in Fig.~\ref{fig: J1J2}.
Both linear and nonlinear conductivities show peak structures around $\omega/|J_1|=1.1$, $1.5$ and $3$.
The peaks at $\omega/|J_1|=1.1$ and $3$ correspond to two magnon resonances at the zone boundary and $q=0$, represented by dashed red and solid red arrows in Fig.~\ref{fig: J1J2}(b), respectively.
The peak at $\omega/|J_1|=1.5$ is the single magnon resonance at $q=0$, represented by a blue arrow in Fig.~\ref{fig: J1J2}(b). 
Single magnon contribution to the conductivities comes from magnons with $q=0$ only and gives a sharp resonance,
while the two magnon contribution arises from magnons with all momenta and shows a broader structure.
It turns out that the single magnon peak in the optical conductivities is relatively larger for larger $S$ due to the factor $S^{3/2}$ in Eq.~\eqref{eq: u0}. 
These results clearly show that the cycloidal phase of $J_1$-$J_2$ spin chain supports shift current at magnon resonances.

One candidate system to observe such shift current response is the cycloidal spin structure in RMnO$_3$. Since both single and two (electro)magnon resonances are observed in the linear optical conductivity in RMnO$_3$ \cite{Aguilar09}, we expect nonzero shift current response induced by those magnon excitations.

\section{Discussions \label{sec: discussion}}

We have demonstrated that magnon shift current generally appears in noncentrosymmetric magnets due to the electric polarization that depends on spin configurations. 
In the present mechanism, dc current flows so as to compensate increasing polarization of electromagnons excited by light irradiation. This increase of polarization is characterized by the factor $1/(\delta \omega+2i\gamma)$ in the nonlinear susceptibility $\chi^{(3)}(\delta \omega; \omega + \delta \omega, -\omega)$ in Eq.~\eqref{eq: chi (3)}, where $\gamma$ represents the dissipation strength for magnons. This factor means that the polarization $P$ increases in time $t$ and is saturated at the relaxation time of magnons $\sim 1/\gamma$. In order to support dc current flow in the steady state, the current (of electrons) should be extracted into the electrodes in a time scale relatively faster than $1/\gamma$.
This indicates that the magnon shift current requires a suitable dissipation mechanism for the underlying electrons that consists of spin systems in the low energy; and an efficient dissipation for electrons (including an efficient coupling to the electrodes) is essential for magnon shift current.

One important question is how large the magnon shift current can be in the present mechanism, especially when compared to the electronic shift current that is induced by optical transition across the electronic band gap.
We can give a crude estimation by comparing the expression for the magnon shift current $\sigma^{(2)}$ in Eq.~\eqref{eq: sigma2} with that for electronic shift current $\sigma^{(2)}_\mathrm{el}$ which is given by \cite{Sipe,Cook17}
\begin{align}
    \sigma^{(2)}_\mathrm{el} =2\pi \int \frac{dk}{2\pi} \sum_{ab} |r_{ab}|^2 R_{ab}f_{ab} \delta(\omega-E_{ab}).
\end{align}
Here, the subscripts $a,b$ label the electronic bands, $r_{ab}=i\braket{a|\partial_k|b}$ is the interband Berry connection with the Bloch wavefunction $\ket{a}$, 
$R_{ab}$ is the shift vector, and $E_{ab}(k)$ is the energy difference between the band $a$ and $b$.  
Thus the ratio between $\sigma^{(2)}$ and $\sigma^{(2)}_\mathrm{el}$ is determined by that between the integrands as
\begin{align}
    \frac{\sigma^{(2)}}{\sigma^{(2)}_\mathrm{el}} 
    &\simeq
    \frac{\widetilde \Pi_{ab}\widetilde \Pi_{ba}(\widetilde \Pi_{aa}-\widetilde \Pi_{bb})}{|r_{ab}|^2 R_{ab}} \frac{E_{ab}}{\epsilon_{ab}},
\end{align}
where $a,b$ are typical states involved in the optical excitations, and we approximated  the delta functions (which leads to the density of states) with the typical magnon and electronic excitation energies (band widths).
For electronic excitations, we adopt typical parameters $r_{ab}\simeq R_{ab}\simeq 0.1$ \AA~  for a ferroelectric material Sn$_2$P$_2$S$_6$ \cite{Sotome-apl} and $E_{ab} \simeq 1$ eV.
For magnon shift current, we consider the superexchange mechanism in Appendix \ref{app: polarization} which gives an estimate of $P$ as
\begin{align}
    P\simeq \frac{aJ}{U} \simeq 10^{-2} \t{\AA},
\end{align}
assuming the energy scale of magnons $J\simeq \epsilon_{ab} \simeq 10$ meV, $U\simeq 3$ eV and $a \simeq 3$ \AA.
These lead to a rough estimation of the ratio as $\sigma^{(2)}/\sigma^{(2)}_\mathrm{el} \simeq 0.1$, indicating that magnon shift current is typically one order smaller than the electronic shift current. We note that the magnon shift current could become larger depending on the magnon energy scale and the involved matrix elements.
Also, the single magnon contribution that we found for the $J_1$-$J_2$ model has the same order of magnitude. One difference is that the single magnon peak is contributed only by $q=0$ state and is typically more significant in the spectrum compared to the two magnon contribution which comes from the magnon continuum and leads to a broad spectrum.
In addition, the single magnon contribution is enhanced for larger $S$
due to the factor $S^{3/2}$ in the magnon-photon coupling.

Finally, we discuss another contribution to the electric polarization in the spin systems. In this paper, we have focused on electronic contribution to the electric polarization. On top of electronic contribution, displacement of ions also contributes to the electric polarization such as magnetostriction mechanism, where ions move depending on the surrounding spin configuration and cause electric polarization. 
For example, we can compare electronic and ionic contributions in the superexchange mechanism for the model with ligand ions in Appendix \ref{app: polarization}. Ionic contribution is estimated by incorporating fluctuation of the ligand ion positions as detailed in Appendix \ref{app: ion}.
The ionic contribution for polarization $P_\mathrm{ion}$ is estimated as
\begin{align}
    P_\mathrm{ion} &= \frac{J}{a M\omega^2} \simeq 10^{-4} \t{\AA},
\end{align}
with the mass of the ion $M$ and the phonon frequency $\omega$, where we consider an oxygen ion as the ligand and the optical phonon $\hbar \omega \simeq 100$ meV. 
This indicates that the ionic contribution $P_\mathrm{ion}$ is usually much smaller than the electronic contribution of the order of $P\simeq 10^{-2}$ \AA. 
Thus the nonlinear optical effects of magnons are also dominated by the coupling between the electronic part of the polarization $P$ and the external electric field, in the case of the superexchange mechanism.

\begin{acknowledgements}
We thank Yoshihiro Okamura, Youtarou Takahashi, Yasuyuki Kato, and Naoto Nagaosa for fruitful discussions.
This work was supported by JST CREST (JPMJCR19T3).
TM acknowledges funding from The University of Tokyo Excellent Young Researcher Program, and JST PRESTO (JPMJPR19L9).
SK acknowledges funding from KAKENHI (20K14407).
\end{acknowledgements}

\appendix

\section{Derivation of Eqs.~(\ref{eq:generic-Heisenberg}), (\ref{eq:polarization-hubbard})\label{app: derivation}}
In this appendix, we provide a detailed derivation of 
\begin{align}
H_{\text{spin}} & =\dfrac{1}{2}\sum_{ij}\dfrac{4|t_{ij}|^{2}\bm{S}_{i}\cdot\bm{S}_{j}}{U-V_{ij}-\bm{E}\cdot\bm{R}_{ij}},\tag{\ref{eq:generic-Heisenberg}}
\end{align}
and  
\begin{align}
\bm{P}_{\text{spin}}&=-\frac{1}{2}\sum_{ij}\dfrac{4|t_{ij}|^{2}\bm{R}_{ij}(\bm{S}_{i}\cdot\bm{S}_{j})}{(U-V_{ij}-\bm{E}\cdot\bm{R}_{ij})^{2}}\tag{\ref{eq:polarization-hubbard-0}}\\
&=-\dfrac{1}{2}\sum_{ij}\dfrac{8UV_{ij}|t_{ij}|^{2}\bm{R}_{ij}(\bm{S}_{i}\cdot\bm{S}_{j})}{(U^{2}-V_{ij}^{2})^{2}}+O(E^2),\tag{\ref{eq:polarization-hubbard}}
\end{align}
 to directly confirm the formula
\begin{equation}
\bm{P}_{\text{spin}}=\hat{P}_0\hat{U}^{\dagger}\bm{P}_\text{el}\hat{U}\hat{P}_0\sim -\dfrac{\partial H_{\text{spin}}}{\partial\bm{E}},\tag{\ref{eq:polarization-formula}}
\end{equation}
which were presented in Sec. \ref{sec: derivation}.
Then we also discuss the deviation of Eq.~(\ref{eq:polarization-formula}) in ac-driven cases.

To obtain Eqs.~(\ref{eq:generic-Heisenberg}) and (\ref{eq:polarization-hubbard}), 
we perform perturbative expansions for the spin Hamiltonian
$H_\text{spin}$ and the polarization operator $\bm{P}_\text{spin}$ as
\begin{align}
    H_\text{spin} &= \sum_n \lambda^n H_\text{spin}^{(n)}, \\
    \bm{P}_\text{spin} &= \sum_n \lambda^n \bm{P}_\text{spin}^{(n)}.
\end{align}
First let us consider the spin Hamiltonian $H_\text{spin}$. The zeroth order term $H_\text{spin}^{(0)}=\hat{P}_0H_\text{loc}\hat{P}_0$ can be regarded as a constant term by definition. The first order term is given as $H_\text{spin}^{(1)}=\hat{P}_0H_\text{hop}\hat{P}_0$, which vanishes upon projection to the spin space as $H_\text{hop}$ changes the number of double occupancy $D$.
The second order reads
\begin{align}
H_\text{spin}^{(2)}&=\hat{P}_0[i\Lambda^{(1)},H_\text{hop}]\hat{P}_0+\hat{P}_0[i\Lambda^{(2)},H_\text{loc}-\bm{E}\cdot\bm{P}_\text{el}]\hat{P}_0\nonumber\\&+\frac{1}{2}\hat{P}_0[i\Lambda^{(1)},[i\Lambda^{(1)},H_\text{loc}-\bm{E}\cdot\bm{P}_\text{el}]]\hat{P}_0
\\
&=\frac{1}{2}\hat{P}_0[i\Lambda^{(1)},H_\text{hop}]\hat{P}_0
\\&=-\frac{1}{2}\hat{P}_0\hat{U}^{(1)}H_\text{hop}\hat{P}_0+\text{H.c.}
\label{eq: Hspin (2)}
\end{align}
Here, the second term in the first line vanishes since $[\hat{P}_0,H_\text{loc}]=[\hat{P}_0,\bm{P}_\text{el}]=0,$ and $\hat{P}_0\Lambda^{(2)}\hat{P}_0=0$.
We have also used Eq.~(\ref{eq:condition-1st-order}) from first to second line.
Using 
\begin{subequations}
\begin{align}
H_{\text{hop}} & =\sum_{ijs}t_{ij}c_{i,s}^{\dagger}c_{j,s},\tag{\ref{eq:generic-Hubbard}a}\\
H_{\text{loc}} & =\sum_{i}[V_{i}(n_{i,\uparrow}+n_{i,\downarrow})+Un_{i,\uparrow}n_{i,\downarrow}],\tag{\ref{eq:generic-Hubbard}b}
\end{align}
\end{subequations}
 we can write Eq.~(\ref{eq:unitary-1st}) as
\begin{equation}
\hat{P}_0\hat{U}^{(1)}=
\sum_{ijs}\dfrac{t_{ij}\hat{P}_0c_{i,s}^{\dagger}c_{j,s}}{U-V_{ij}-\bm{E}\cdot\bm{R}_{ij}},
\end{equation}
where $\bm{R}_{ij}=\bm{R}_i-\bm{R}_{j}, V_{ij}=V_i-V_j$.
Using this expression, we obtain
\begin{align}
&\hat{P}_0\hat{U}^{(1)}H_\text{hop}\hat{P}_0 \nonumber \\
&=
\sum_{ijs}\dfrac{t_{ij}\hat{P}_0c_{i,s}^{\dagger}c_{j,s}}{U-V_{ij}-\bm{E}\cdot\bm{R}_{ij}}\sum_{s^\prime}t_{ji}c_{j,s^\prime}^\dagger c_{i,s^\prime}\hat{P}_0\\
&=\sum_{ijss^\prime}\dfrac{|t_{ij}|^2\left(\delta_{s,s^\prime}-4\bm{S}_i\cdot\bm{\sigma}_{s^\prime s}\bm{S}_j\cdot\bm{\sigma}_{s s^\prime}\right)}{4(U-V_{ij}-\bm{E}\cdot\bm{R}_{ij})} \\
&=-2 \sum_{ij}
\dfrac{|t_{ij}|^2 \bm{S}_i\cdot \bm{S}_j}{U-V_{ij}-\bm{E}\cdot\bm{R}_{ij}} + \t{const.}.
\end{align}
Here we have used 
$\hat{P}_0c_{i,s}^\dagger c_{i,s^\prime}\hat{P}_0=\delta_{s,s^\prime}/2+\bm{S}_i\cdot\bm{\sigma}_{s^\prime s}$ 
with the Pauli matrices $\sigma_i$, and 
$\t{tr}[\sigma_i \sigma_j]=2 \delta_{i,j}$.
Once we plug in the above equation to Eq.~\eqref{eq: Hspin (2)}, we end up with the spin Hamiltonian (\ref{eq:generic-Heisenberg}).

The effective polarization operator can be directly evaluated in a similar manner. 
The perturbative evaluation reads
\begin{align}
    \bm{P}_{\text{spin}}^{(2)}&=\hat{P}_0[i\Lambda^{(2)},\bm{P}_\text{el}]\hat{P}_0+\frac{1}{2}\hat{P}_0[i\Lambda^{(1)},[i\Lambda^{(1)},\bm{P}_\text{el}]]\hat{P}_0\\
    &=\frac{1}{2}\hat{P}_0 i\Lambda^{(1)} [i\Lambda^{(1)},\bm{P}_\text{el}]\hat{P}_0 
    - \frac{1}{2}\hat{P}_0 [i\Lambda^{(1)},\bm{P}_\text{el}] i\Lambda^{(1)} \hat{P}_0 \\
&=-\frac{1}{2}\hat{P}_0i\Lambda^{(1)}[\bm{P}_\text{el},i\Lambda^{(1)}\hat{P}_0]+\text{H.c.}\\
    &=\frac{1}{2}
\hat{P}_0\hat{U}^{(1)}[\bm{P}_\text{el},(\hat{P}_0\hat{U}^{(1)})^\dagger]+\text{H.c.}
\end{align}
where the term with $\Lambda^{(2)}$ in the first line vanishes in a similar manner as for $H_\text{spin}^{(2)}$, and we used $[\hat{P}_0,\bm{P}_\text{el}]=0$ from the second line to the third line.
Using the expression for $\hat{P}_0\hat{U}^{(1)}$, we obtain
\begin{align}
[\bm{P}_\text{el},(\hat{P}_0\hat{U}^{(1)})^\dagger]&=\sum_{ijs}\dfrac{\bm{R}_{ij}t_{ji}c_{j,s}^{\dagger}c_{i,s}\hat{P}_0}{U-V_{ij}-\bm{E}\cdot\bm{R}_{ij}},
\end{align}
which leads to Eq.~(\ref{eq:polarization-hubbard}).
Namely, Eq.~(\ref{eq:polarization-formula}) holds within the second-order perturbation in $\lambda$, including higher orders of $E$.

We note that the formula (\ref{eq:polarization-formula}) slightly deviates when we consider ac electric
fields. Control of spin systems with ac electric fields are recently studied actively in the context of Floquet engineering \cite{Oka-Kitamura,Mentink2014,Itin2015,Bukov2016,Kitamura17,Claassen2017,Mentink2017,Losada19}, where the effective spin Hamiltonian is obtained by making $\hat{U}$ time-dependent.
In the time-dependent case, we have an additional term
as
\begin{equation}
H_{\text{eff}}=\hat{U}^{\dagger}(t)(H_{\text{el}}-\bm{E}(t)\cdot\bm{P}_\text{el})\hat{U}(t)-i\hat{U}^{\dagger}(t)\dfrac{d}{dt}\hat{U}(t),
\end{equation}
which modifies Eq.~(\ref{eq:polarization-derivation}) into 
\begin{align}
\dfrac{\partial H_{\text{spin}}}{\partial\bm{E}(t)} & =-\bm{P}_\text{spin}-i\dfrac{d}{dt}\left(\hat{P}_0\hat{U}^{\dagger}\partial_{\bm{E}}\hat{U}\hat{P}_0\right)\nonumber\\&+\left[H_{\text{spin}},\hat{P}_0\hat{U}^{\dagger}\partial_{\bm{E}}\hat{U}\hat{P}_0\right].
\end{align}
The second term may appear in $\mathcal{O}(\lambda^{2})$ which is the same order with the first term, and thus Eq.~(\ref{eq:polarization-formula})
does not hold in general. However, we can neglect the additional term
in several situations. One is when the driving frequency $\omega$ is so high
that we can replace the expression by the time-averaged one. In this case, a sufficiently slow (dc) component of $P$ is contributed by the second term of the order of $1/T$ with the (long) period $T$ which becomes negligible. Another
is when the driving frequency $\omega\sim(d/dt)$ is sufficiently
small. Indeed, we can check these by combining the Floquet theory with Schrieffer-Wolff transformation, which results in a spin Hamiltonian with time-periodic coupling [See Ref.~\cite{Kitamura17}].
For the Hamiltonian given by Eq.~(\ref{eq:generic-Hubbard}), we obtain
\begin{align}
\dfrac{\partial H_{\text{spin}}}{\partial\bm{E}(t)} & \sim\dfrac{1}{2}\sum_{ij}\dfrac{4|t_{ij}|^{2}\bm{R}_{ij}(\bm{S}_{i}\cdot\bm{S}_{j})}{(U-V_{ij})^{2}-\omega^{2}},
\label{eq: P RMH app}
\end{align}
in the leading order of $E$. Namely, as
we are interested in optical excitation of magnons, we can neglect
the correction $\sim\omega^{2}/U^{2}\ll1$, which 
reproduces  Eq.~\eqref{eq:polarization-hubbard}.

\section{Polarization in the superexchange mechanism \label{app: polarization}}

\begin{figure}
\begin{centering}
\includegraphics[width=1\columnwidth]{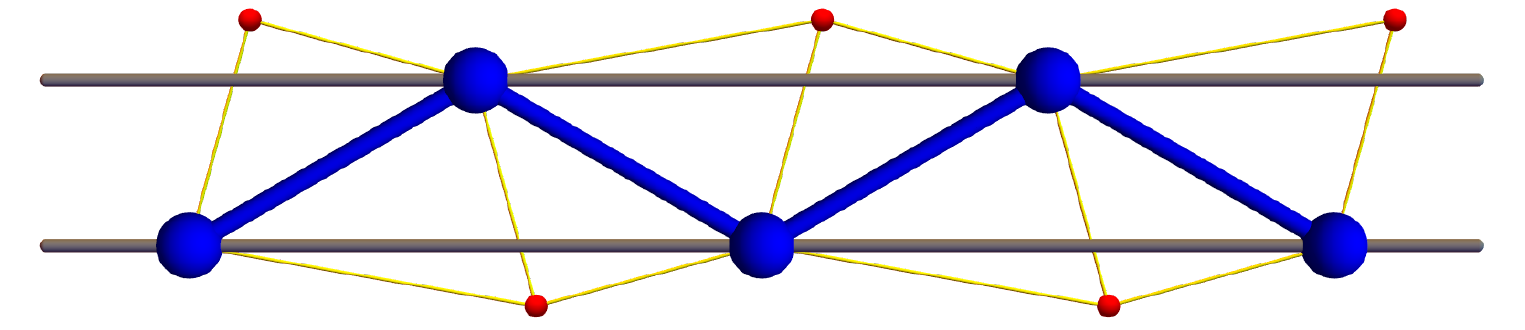}\caption{\label{fig:ligand}An example of configuration with nonzero polarization operator. The
blue spheres represent magnetic ions, while the red ones are ligand
ions. Electrons can virtually hop along the yellow bonds, which leads
to the superexchange of spins on the magnetic sites, which is represented
by blue and gray bonds.}
\par\end{centering}
\end{figure}

In this section, we study electric polarization induced by the superexchange mechanism, 
where the spin sites are connected via ligand ions as illustrated in Fig.~\ref{fig:ligand}.

The expression
for the effective polarization operator Eq.~(\ref{eq:polarization-hubbard}) implies that
the electric polarization is induced along the potential difference
on the path of the kinetic exchange process. We can apply this idea
to the superexchange mechanism via the ligand ions as well, where
the potential difference naturally appears. While the potential difference
between magnetic sites cannot be as large as $U$ with keeping the
half-filled situation, the potential difference with the ligand sites
may be, which can be utilized for enhancing the current response.

We here consider a Hubbard model consisting of two magnetic ions $d_{js}^{\dagger}$
at $\bm{R}_{j}$ with $j=1,2$ and one ligand ion $p_{m\sigma}^{\dagger}$
at $\bm{R}_{p}$ with $m=x,y$, whose Hamiltonian is given as $H=H_{\text{hop}}+H_{\text{loc}}$, 
\begin{subequations}
\begin{align}
H_{\text{hop}} & =-\sum_{s}(t_{1}d_{1,s}^{\dagger}p_{x,s}+t_{2}d_{2,s}^{\dagger}p_{\theta,s})+\text{H.c.},\label{eq:dpHamiltonian}\\
H_{\text{loc}} & =\dfrac{1}{2}U_{d}\sum_{j}(n_{j}^{d}-1)^{2}+\dfrac{1}{2}U_{p}(n^{p}-4)^{2}\nonumber \\
 & -J_{H}\bm{S}^{p}\cdot\bm{S}^{p}+\sum_{j}V_{j}^{d}n_{j}^{d}+V^{p}n^{p},
\end{align}
\end{subequations}
where $p_{\theta,s}=\cos\theta p_{x,s}+\sin\theta p_{y,s}$ with $\theta=\angle\bm{R}_{1}\bm{R}_{p}\bm{R}_{2}$,
$n_{j}^{d}=\sum_{s}d_{j,s}^{\dagger}d_{j,s},n^{p}=\sum_{m,s}p_{m,s}^{\dagger}p_{m,s},$
and
\begin{equation}
\bm{S}^{p}=\sum_{m=x,y}\sum_{s,s^{\prime}}\dfrac{1}{2}p_{m,s}^{\dagger}\bm{\sigma}_{s,s^{\prime}}p_{m,s^{\prime}}.
\end{equation}
We assume that $n_{j}^{d}=1,n^{p}=4$ is realized in the atomic limit.
Note that the onsite interaction for $p$ is rotationally invariant,
and thus we have set $p_{x}$ as the orbital parallel to $\bm{R}_{1p}=\bm{R}_{1}-\bm{R}_{p}$. 

The superexchange process under the applied dc electric field is studied
in Ref.~\cite{Furuya21}. Following the
calculation there, the spin Hamiltonian for magnetic ions is obtained
as
\begin{align}
H_{\text{spin}} & =J_{12}\bm{S}_{1}\cdot\bm{S}_{2} \\
 & =(J_{A}\cos^{2}\theta-J_{F}\sin^{2}\theta)\bm{S}_{1}\cdot\bm{S}_{2},\label{eq:superexchange-theta}
\end{align}
where
\begin{subequations}
\begin{align}
J_{A} & =\dfrac{2|t_{1}t_{2}|^{2}}{(U_{d}+\Delta_{1}-\Delta_{2})\Delta_{1}^{2}}+\dfrac{2|t_{1}t_{2}|^{2}}{(U_{d}+\Delta_{2}-\Delta_{1})\Delta_{2}^{2}}\nonumber \\
 & +\left(\dfrac{1}{\Delta_{1}}+\dfrac{1}{\Delta_{2}}\right)^{2}\dfrac{2|t_{1}t_{2}|^{2}}{2\Delta_{H}+J_{H}},\\
J_{F} & =\left(\dfrac{1}{\Delta_{1}}+\dfrac{1}{\Delta_{2}}\right)^{2}\dfrac{2|t_{1}t_{2}|^{2}J_{H}}{4\Delta_{H}^{2}-J_{H}^{2}}
\end{align}\label{eq:superexchange}
\end{subequations}
describe antiferromagnetic and ferromagnetic interaction, respectively,
and 
\begin{align}
\Delta_{j} & =\dfrac{1}{2}(U_{d}+U_{p})+V_{jp}-\dfrac{3}{4}J_{H},\\
\Delta_{H} & =\dfrac{1}{2}(U_{d}+2U_{p}+V_{1p}+V_{2p}-J_{H}).
\end{align}

The effective polarization operator can be obtained by setting $V_{jp}=V_{j}^{d}-V^{p}\rightarrow V_{jp}+\bm{E}\cdot\bm{R}_{jp}$
and taking derivative. In the leading order of $E$, we obtain $H_{\text{spin}}=(J_{12}+\bm{E}\cdot\bm{\Pi}_{12})\bm{S}_{1}\cdot\bm{S}_{2}$
with 
\begin{align}
\bm{\Pi}_{12} & =\left(\dfrac{\bm{R}_{1}+\bm{R}_{2}}{2}-\bm{R}_{p}\right)\left(\partial_{\Delta_{1}}+\partial_{\Delta_{2}}+\partial_{\Delta_{H}}\right)J_{12}\nonumber \\
 & +\dfrac{1}{2}\bm{R}_{12}\left(\partial_{\Delta_{1}}-\partial_{\Delta_{2}}\right)J_{12}.
\end{align}
Here, the second term is the counterpart of Eq.~(\ref{eq:polarization-hubbard}), i.e.,
the polarization induced by the potential difference between magnetic
sites $\Delta_{1}-\Delta_{2}=V_{1}-V_{2}$, which is evaluated as
\begin{subequations}
\begin{align}
\left(\partial_{\Delta_{1}}-\partial_{\Delta_{2}}\right)J_{A} & \sim\dfrac{32|t_{1}t_{2}|^{2}(V_{1}-V_{2})}{(U_{d}+2V_{1p})^{5}}\nonumber \\
 & \times\left(4-\dfrac{3V_{1p}}{U_{d}}+\dfrac{12V_{1p}^{2}}{U_{d}^{2}}+\dfrac{4V_{1p}^{3}}{U_{d}^{3}}\right),\\
\left(\partial_{\Delta_{1}}-\partial_{\Delta_{2}}\right)J_{F} & \sim\dfrac{256|t_{1}t_{2}|^{2}(V_{1}-V_{2})J_{H}}{(U_{d}+2V_{1p})^{6}},
\end{align}
\end{subequations}
where we have neglect $U_{p}$ and the higher order of $J_{H},V_{1}-V_{2}$
for simplicity. On the other hand, the first term is evaluated under
the same condition as
\begin{subequations}
\begin{gather}
\left(\partial_{\Delta_{1}}+\partial_{\Delta_{2}}+\partial_{\Delta_{H}}\right)J_{A}\sim-\dfrac{128|t_{1}t_{2}|^{2}}{(U_{d}+2V_{1p})^{4}}\left(2+\dfrac{V_{1p}}{U_{d}}\right),\\
\left(\partial_{\Delta_{1}}+\partial_{\Delta_{2}}+\partial_{\Delta_{H}}\right)J_{F}\sim-\dfrac{256|t_{1}t_{2}|^{2}J_{H}}{(U_{d}+2V_{1p})^{5}},
\end{gather}
\end{subequations}
which appear due to the potential difference between magnetic and
ligand ions, and may be much larger. 


Note that the first term vanishes when the bond center coincides with
the position of the ligand ion (or the average over equidistant ions).
Here, let us provide an example where the first term does not vanish. 
As depicted in Fig.~\ref{fig:ligand}, we put the magnetic ions in
a zigzag geometry as
\begin{equation}
\bm{R}_{j}=(2j\cos\phi,(-1)^{j}\sin\phi),
\end{equation}
with $j\in\mathbb{Z}$, $0<\phi<\pi/4$, while the ligand ions are
placed at
\begin{equation}
\bm{R}_{j}^{p}=((2j+1)\cos\phi-\sin\phi,(-1)^{j+1}\cos\phi),
\end{equation}
where $\angle\bm{R}_{j}\bm{R}_{j}^{p}\bm{R}_{j+1}=\pi/2$. With this
configuration, we obtain ferromagnetic and antiferromagnetic exchange
interactions for nearest and next-nearest neighbors, respectively.
We obtain nonzero $\bm{\Pi}_{ij}$ for these interactions via
\begin{subequations}
\begin{align}
\dfrac{\bm{R}_{j}+\bm{R}_{j+1}}{2}-\bm{R}_{j}^{p} & =(\sin\phi,(-1)^{j}\cos\phi),\label{eq:superexchange-polarization}\\
\dfrac{\bm{R}_{j}+\bm{R}_{j+2}}{2}-\bm{R}_{j+1}^{p} & =\dfrac{1}{\sqrt{2}}\sin\left(\phi-\dfrac{\pi}{4}\right)(1,(-1)^{j}),
\end{align}
\end{subequations}
if only the shortest hopping path is taken into account. 
If we focus on the transport along e.g., $(\cos\phi,\sin\phi)$ direction, Eq.~(\ref{eq:superexchange-polarization}) describes polarization operator of a form (\ref{eq:P-J1J2}) with $p_1=\delta p_1$.

\section{Magnon representation and optical conductivity for $J_1$-$J_2$ spin chains \label{app: j1j2}}
In this section, we derive the magnon representation of a bilinear form of spin operators in the cycloidal phase of $J_1$-$J_2$ spin chains. 
This representation is applicable to the magnon Hamiltonian and the polarization operator.

We consider the cycloidal spin structure for the $J_1$-$J_2$ spin chain defined by Eq.~(\ref{eq: H J1J2}) in the main text.
We rotate the spin quantization axis along the cycloidal spin structure in Eq.~(\ref{eq: cycloidal}) as $\bm{S}_{l}=\mathcal{R}\tilde{\bm{S}}_{l}$, where $\mathcal{R}$ is the rotation matrix given by
\begin{align}
	\mathcal{R} = \left(
	\begin{array}{c c c}
		\cos{Qla} & 0 & \sin{Qla} \\
		0 & 1 & 0 \\
		-\sin{Qla} & 0 & \cos{Qla}
	\end{array}
	\right),
\end{align} 
and $\tilde{\bm{S}}_{l}$ is the spin structure in the rotated spin coordinate.
For example, a Heisenberg term $\bm S_i \cdot \bm S_{i+n}$ is expressed with the rotated spin coordinate as
\begin{align}
    \bm S_i \cdot \bm S_{i+n}&=
    \tilde S_i^z (\tilde S_{i+n}^z \cos nQa - \tilde S_{i+n}^x \sin nQa) \n
    &+\tilde S_i^x (\tilde S_{i+n}^z \sin nQa + \tilde S_{i+n}^x \cos nQa) \n
    &+\tilde S_i^y \tilde S_{i+n}^y.
\end{align}
We introduce the magnon excitations using Holstein-Primakoff transformation in Eq.~(\ref{eq: HP transf}) in the rotated spin coordinate, which reads
\begin{align}
    \tilde S^x_i&=\sqrt{\frac{S}{2}}(a_i^\dagger+a_i), \\
    \tilde S^y_i&=-i\sqrt{\frac{S}{2}}(a_i^\dagger-a_i), \\
    \tilde S^z_i&=S-a_i^\dagger a_i, 
\end{align}
assuming large $S$ and small magnon density.

\subsection{Magnon representation for Heisenberg interactions}
Now we derive the magnon representation for a bilinear form of spin operators, 
\begin{align}
\mathcal{O}&=
\sum_{n=1,2}\sum_i c_{i,n} \bm{S}_i \cdot \bm{S}_{i+n}.
\label{eq: O cin}
\end{align}
Here we consider Heisenberg interactions $c_{i,n}$ between the $i$th spin and the $(i+n)$th spin, up to $n=2$ (next nearest neighbors).
By expanding Eq.~(\ref{eq: O cin}) with respect to the second-order of $a^{\;}_{i}$ and $a^\dagger_{i}$, we obtain 
\begin{align}
\mathcal{O} &= \sum_{n=1,2} \sum_i c_{i,n} S  \n
&\times \Bigg\{ \cos^2\frac{nQa}{2}\left(a^\dagger_{i}a^{\;}_{i+n}+a^{\;}_{i}a^\dagger_{i+n}\right)\nonumber\\
&- \sin^2\frac{nQa}{2}\left(a^{\;}_{i}a^{\;}_{i+n}+a^\dagger_{i}a^\dagger_{i+n}\right)\nonumber\\
&- \cos{nQa}\left(a^\dagger_{i}a^{\;}_{i}+a^{\;}_{i+n}a^\dagger_{i+n}\right)\nonumber\\
&+\sqrt{\frac{S}{2}}\sin{nQa}\left(a^\dagger_{i}+a^{\;}_{i}-a^{\;}_{i+n}-a^\dagger_{i+n}\right)\Bigg\}
\label{eq: O cin a}
\end{align}
up to constant.

Hereafter, we consider the sublattice $A$ and $B$ for even and odd sites, respectively.
In addition to uniform Heisenberg interactions $c_1$ and $c_2$ between nearest neighbors and next nearest neighbors, we introduce alternation for the Heisenberg interaction  $\delta c_1$ between nearest neighbors.
Specifically, we consider the bilinear form of spin operators
\begin{align}
\mathcal{O}&=
\sum_i \{c_1 + \delta c_1 (-1)^i\} \bm{S}_i \cdot \bm{S}_{i+1} 
+ c_2 \sum_i \bm{S}_i \cdot \bm{S}_{i+2}.
\end{align}
By substituting $c_{i,1}=c_1+(-1)^i\delta c_1$ and $c_{i,2}=c_2$ in Eq.~\eqref{eq: O cin a} and performing Fourier transformation, we obtain 
\begin{align}
\mathcal{O}&=
\sum_q
\Psi_q^\dagger
\begin{pmatrix}
A_q &  B_q & C_{-q} &  D_{-q} \\
B_q &  A_q & D_{-q} &  C_{-q} \\
C_{q} &  D_{q} & A_q &  B_q  \\
D_{q} &  C_{q} & B_q &  A_q  \\
\end{pmatrix}
\Psi_q 
+ u_0\Psi_{q=0}, 
\label{eq: O magnon rep}
\end{align}
with
\begin{align}
\Psi_q &=
\begin{pmatrix}
a_{qA} \\  a_{-qA}^\dagger \\ a_{qB} \\ a_{-qB}^\dagger
\end{pmatrix},
\end{align}
where $a_{qA(B)}$ is the annihilation operator of magnon with the momentum $q$ on the sublattice A(B).
The coefficients are given by
\begin{align}
A_q &=-c_1S \cos Qa + c_2S \{\cos^2 Qa \cos 2qa - \cos 2Qa \}, \\
B_q &= -c_2S \sin^2Qa \cos 2qa, \\
C_q &= S\cos^2\frac{Qa}{2}(c_1 \cos qa +i\delta c_1 \sin qa), \\
D_q &= -S\sin^2\frac{Qa}{2}(c_1 \cos qa +i\delta c_1 \sin qa), \\
u_0 &= 2S\sqrt{\frac{S}{2}}\delta c_1\sin{Qa} (1,1,-1,-1).
\end{align}
Note that the $q$ summation in Eq.~\eqref{eq: O magnon rep} runs over the whole Brillouin zone and hence $\Psi_q$ covers each mode ($a_{qA/B}$) twice in the BdG representation.

Since the Hamiltonian \eqref{eq: H J1J2} has only uniform Heisenberg couplings, its magnon representation is given by bilinear form of magnon operators (i.e., $u_0=0$ in Eq.~\eqref{eq: O magnon rep}).
In contrast, the polarization operator $P$ contains alternation $\delta \Pi_1$ and has a single magnon term in its magnon representation. 
This term describes the (staggered) tilting of the quantization axis
from the cycloidal spin structure, since the term can be absorbed
by introducing new bosonic operators $\tilde{a}_{q=0,A/B}=a_{q=0,A/B}\mp\sqrt{S/2}(\delta c_{1}/2c_{1})\tan Qa$,
which implies that the eigenstate of $\mathcal{O}$ satisfies $\sqrt{2S}\langle a\rangle=\langle\tilde{S}^{x}+i\tilde{S}^{y}\rangle\neq0$.
This single magnon terms enables a direct coupling of a single magnon excitation to an external electric field, which gives rise to single magnon resoances in the optical conductivities.
While the two magnon excitation from the first term in Eq.~\eqref{eq: O magnon rep} can be treated with the formulation presented in Sec.~\ref{sec: conductivities},
this single magnon resonance needs to be treated separately as we describe below.
 
\subsection{Single magnon contribution to the optical conductivities}
Next let us consider the contributions of the single magnon excitations to the optical conductivities in the $J_1$-$J_2$ spin chains.
Suppose that the electric polarization $P$ is written in terms of magnon operators as
\begin{align}
    P&=-\left(\sum_q \Psi_q^\dagger \Pi_q \Psi_q + \pi_0 \Psi_{q=0}\right).
\end{align}
The second term proportional to $\pi_0$ (which is a row vector) gives rise to the single magnon contribution to the optical conductivities. 
In the following, we consider the contribution from $\pi_0$ to the conductivities.

First, the single magnon contribution to the linear conductivity $\sigma^{(1)}_{1ph}$ is given by 
\begin{align}
    \sigma^{(1)}_{1ph}(\omega)&=-\frac{i\omega}{V_\t{cell}} (-\pi_0) G(\omega,q=0) \pi_0^T.
\end{align}
Here, $V_\t{cell}$ is the unit cell volume and $V_\t{cell}=2a$ with the lattice constant $a$ for our representation \eqref{eq: O magnon rep} for the $J_1$-$J_2$ spin chains.
Using the expression $G(\omega,q)=V(\omega-E_q + i\gamma)^{-1}V^{-1}B$ in Sec.~\ref{sec: conductivities},
we can write
\begin{align}
    \sigma^{(1)}_{1ph}(\omega)&=\frac{i\omega}{V_\t{cell}} \frac{(\pi_0 V)_a (V^{-1}B \pi_0^T)_a}{\omega-\epsilon_a+i\gamma},
\end{align}
where the subscripts $a$ in the numerator denotes the $a$th component of the vectors. 

Next the single magnon contribution to the nonlinear conductivity $\sigma^{(2)}_{1ph}$ is given by
\begin{align}
    \sigma^{(2)}_{1ph}(\omega)&=\lim_{\delta \omega \to 0} (-i\delta\omega)\chi^{(3)}_{1ph}(\delta\omega;\omega+\delta\omega, -\omega).
\end{align}
Here the single magnon contribution to the nonlinear susceptibility $\chi^{(3)}_{1ph}$ is defined as
\begin{align}
    &\chi^{(3)}_{1ph}(i\Omega_1+i\Omega_2;i\Omega_1,i\Omega_2) \n
    &= \frac{1}{V_\t{cell}}
    \Big[\pi_0 G(-i\Omega_2,q=0) (-\Pi_{q=0}) G(i\Omega_1,q=0) \pi_0^T \n
    &\qquad +\pi_0 G(-i\Omega_1,q=0) (-\Pi_{q=0}) G(i\Omega_2,q=0) \pi_0^T
    \Big].
\end{align}
In this expression, two incoming photons couples with the magnons with the single magnon term $\pi_0$, and the resulting polarization (that causes shift current response) is induced by the two magnon term $\Pi_{q=0}$.
(In principle, $\chi^{(3)}$ has an additional contribution from an $E^2$ term in the polarization operator $P$ which we neglected. This contribution does not lead to $1/\delta\omega$ divergence in $\chi^{(3)}$ which justifies the present treatment.)
Using $G(\omega,q)=V(\omega-E_q + i\gamma)^{-1}V^{-1}B$, we can rewrite $\chi^{(3)}_{1ph}$ as
\begin{align}
    &V_\t{cell}\chi^{(3)}_{1ph}(i\Omega_1+i\Omega_2;i\Omega_1,i\Omega_2) \n
    &= -\pi_0 V(-i\Omega_2-E_q)^{-1}V^{-1}B \Pi_{q=0} V(i\Omega_1-E_q)^{-1}V^{-1}B \pi_0^T \n
    &\qquad +(i\Omega_1 \leftrightarrow i\Omega_2) \n
    &= -\sum_{ab}  \frac{(\pi_0 V)_a (\widetilde \Pi_{q=0})_{ab} (V^{-1}B \pi_0^T)_b}{(-i\Omega_2 - \epsilon_a)(i\Omega_1 - \epsilon_b)}
    +(i\Omega_1 \leftrightarrow i\Omega_2).
\end{align}
The divergent term $\propto 1/\delta\omega$ originates from the term with $a=b$, as can be seen from
\begin{align}
    &\frac{1}{(-i\Omega_2 - \epsilon_a)(i\Omega_1 - \epsilon_a)} \n
    &=\frac{1}{i\Omega_1+i\Omega_2}
    \left(
    \frac{1}{-i\Omega_2 - \epsilon_a} - \frac{1}{i\Omega_1 - \epsilon_a}
    \right)
\end{align}
Performing the analytic continuation in Eq.~\eqref{eq: analytic continuation}, we obtain
\begin{align}
    &\lim_{\delta\omega \to 0} \delta\omega \chi^{(3)}_{1ph}(\delta\omega;\omega+\delta\omega, -\omega) \n
    &= -\frac{1}{V_\t{cell}}\sum_{a} (\pi_0 V)_a (\widetilde \Pi_{q=0})_{aa} (V^{-1}B \pi_0^T)_a \n
    &\times \Bigg[\frac{1}{(\omega - \epsilon_a -i\gamma)}-\frac{1}{(\omega - \epsilon_a +i\gamma)} 
    + (\omega \leftrightarrow -\omega) \Bigg]
\end{align}
Therefore, we obtain $\sigma^{(2)}_{1ph}$ as
\begin{align}
    \sigma^{(2)}_{1ph}(\omega) &= -\frac{2\pi}{V_\t{cell}} \sum_a 
    (\pi_0 V)_a (\widetilde \Pi_{q=0})_{aa} (V^{-1}B \pi_0^T)_a \n
    & \qquad \times [\delta(\omega - \epsilon_a) + \delta(-\omega - \epsilon_a)].
\end{align}
This expression clearly indicates that the polarization $(\widetilde{\Pi}_{q=0})_{aa}$ of the magnon in the $a$th branch induces dc current upon its optical excitation.

\section{Comparison of electronic and ionic contributions to the electric polarization \label{app: ion}}
In this Appendix, we compare electronic and ionic contributions to the electric polarization in the superexchange mechanism discussed in Appendix \ref{app: polarization}.
Here, let us roughly estimate the electric polarization induced by the fluctuation
of the ligand ion position. 
Such effect can be treated by introducing the positional fluctuation as $\bm{R}_p\rightarrow\bm{R}_p+\bm{\delta}$ [in Eq.~(\ref{eq:dpHamiltonian})] and 
regarding
$\bm{\delta}$ as dynamical degrees of freedom, i.e., phonon.
The coupling between the atomic displacement and spin is then encoded
via the modulation of the exchange interaction. 

When the positions of ligand ions are modulated, they affect the exchange coupling via (i) change in the bond angle and (ii) change in the hopping amplitude due to the modulated bond length. 
Let us take into account the modulation of the exchange coupling, $\delta J$, up to the first order in $\delta$.
For the former contribution, we obtain 
\begin{align}
\delta J & =-2(J_{A}-J_{F})\dfrac{\bm{R}_{1p}\cdot\bm{R}_{2p}}{|\bm{R}_{1p}|^{2}|\bm{R}_{2p}|^{2}}\nonumber\\&\times\left[\dfrac{\bm{R}_{1p}}{|\bm{R}_{1p}|^{2}}(\bm{R}_{1p}\cdot\bm{R}_{12})-\dfrac{\bm{R}_{2p}}{|\bm{R}_{2p}|^{2}}(\bm{R}_{2p}\cdot\bm{R}_{12})\right]\cdot\bm{\delta},
\end{align}
by inserting the modulation of the bond angle $\theta=\angle\bm{R}_{1}\bm{R}_{p}\bm{R}_{2}$ to Eq.~(\ref{eq:superexchange-theta}).
Namely, $\delta J\sim J \delta/a$ with $a$ being the lattice constant. 
For the latter contribution, we assume that the hopping amplitude scales as  $t_{ij}=t_{0}e^{-|\bm{R}_{i}-\bm{R}_{j}|/\xi}$, which leads to 
\begin{equation}
\delta J=\dfrac{2J}{\xi}\left(\dfrac{\bm{R}_{1p}}{|\bm{R}_{1p}|}+\dfrac{\bm{R}_{2p}}{|\bm{R}_{2p}|}\right)\cdot\bm{\delta}
\end{equation} 
via the modulation of $|t_1t_2|^2$ in Eq.~(\ref{eq:superexchange}). This contribution $\delta J\sim J\delta/\xi$ may be comparable to the former one when $\xi\sim a$.

The ionic contribution to the exchange striction emerges when the expectation value of $\bm{\delta}$ is perturbed by the applied
electric field. Namely, we obtain the field-induced modulation of the exchange  interaction $J\rightarrow J+\bm{E}\cdot\bm{\Pi}$ as 
\begin{equation}
\bm{\Pi}=\dfrac{\partial(\delta J)}{\partial\bm{\delta}}\cdot\dfrac{\partial\langle\bm{\delta}\rangle}{\partial\bm{E}}.
\end{equation}
Here let us roughly estimate the size of this effect. As we have seen, $\partial_\delta(\delta J)\sim J/a$ (or $\sim J/\xi$). We estimate $|\partial_{\bm{E}}\langle\bm{\delta}\rangle|$ as follows. We model the phonon Hamiltonian by a harmonic
oscillator as $H_{\text{ph}}=\bm{P}^2/(2M)+M\omega^2\bm{\delta}^2/2$, where $\bm{P}$ is the momentum conjugate to $\bm{\delta}$, and $M$ is the mass of nuclei. The characteristic frequency $\omega$ should correspond to that of optical phonons in more realistic descriptions. When the electric field is applied, we obtain 
\begin{align}
H_{\text{ph}}-q\bm{E}\cdot\bm{\delta} &=\dfrac{\bm{P}^{2}}{2M}+\dfrac{1}{2}M\omega^{2}\left(\bm{\delta}-\dfrac{q\bm{E}}{M\omega^{2}}\right)^{2}-\dfrac{q^{2}\bm{E}^{2}}{2M\omega^{2}},
\end{align}
which implies that the position of the oscillator should shift by $q\bm{E}(M\omega^{2})^{-1}$
in the adiabatic picture ($q$ is the total charge). 
Since $(M\omega^{2})^{-1}\sim(3800(\hbar\omega)^{2})^{-1}\text{eV}\cdot\textrm{\AA}^{2}$
for the oxygen atom, we obtain an estimation $|\partial_{\bm{E}}\langle\bm{\delta}\rangle|\sim(M\omega^{2})^{-1}\sim(1/38)\textrm{\AA}^{2}/\text{eV}$
for an optical phonon with $\hbar\omega\sim100\text{meV}$.
Therefore, the ionic contribution to the elctric polarization can be estimated as
\begin{align}
    \Pi \simeq \frac{J}{a M\omega^2} \simeq 10^{-4} \t{\AA},
\end{align}
for $J\simeq 10$ meV and $a \simeq 3$ \AA.
This should be compared with the electronic contribution $\Pi\simeq a J/U \simeq 10^{-2} \t{\AA}$ for $U=3$ eV and suggests that the ionic contribution is usually much smaller than the electronic contribution in the superexhcange mechanism.

\bibliography{reference}

\end{document}